\documentclass[12pt]{elsart}
\usepackage{epsfig,graphics}
\newlength{\lslash}
\def\slash#1{\settowidth{\lslash}{$#1$}\makebox[\lslash]{\makebox[0mm]{$/$}\makebox[0mm]{$#1$}}}
\newcommand{\tr}{\rm tr \,}

\newcommand{\chiral}[1]{\stackrel{\circ}{#1}}
\def\Re{{\rm Re\,}}
\def\Im{{\rm Im\,}}

\newcommand{\TableHeader}[0]{\begin{tabular}{|c|c|c|c|c|c|} \hline (I,S)  &  Kanal &  Lage &  |g|  & Lage  &|g| \\}
\def\rescale{\fontsize{7}{2}}
\begin{document}
\begin{frontmatter}
\title{Baryon self energies in the chiral loop expansion }
\author[GSI]{A. Semke}
\author[GSI]{and M.F.M. Lutz}
\address[GSI]{Gesellschaft f\"ur Schwerionenforschung (GSI)\\
Planck Str. 1, 64291 Darmstadt, Germany}
\begin{abstract}
We compute the self energies of the baryon octet and decuplet states at the one-loop level applying
the manifestly covariant chiral Lagrangian. It is demonstrated that expressions consistent
with the expectation of power counting rules arise if the self energies are decomposed according
to the Passarino-Veltman scheme supplemented by a minimal subtraction. This
defines a partial summation of the chiral expansion. A finite renormalization
required to install chiral power counting rules leads to the presence of an infrared renormalization
scale. Good convergence properties for the chiral loop expansion of the baryon octet and decuplet masses
are obtained for natural values of the infrared scale. A prediction for the strange-quark matrix element of
the nucleon is made.
\end{abstract}
\end{frontmatter}

\section{Introduction}

The application of chiral perturbation theory to the SU(3) flavor sector of QCD is hampered
by poor convergence properties for processes involving
baryons \cite{Jenkins:1992,Bernard:Kaiser:Meissner:1993,Borasoy:Meissner:1997,Ellis:Torikoshi:1999,Lehnhart:Gegelia:Scherer:2005}. The
original computation of the nucleon self energy by Gasser, Sainio and Svarc \cite{Gasser:Sainio:Svarc:1988}
was performed as an application of the manifestly covariant chiral Lagrangian. It was observed that
the $\overline{MS}$ scheme \cite{Gasser:Leutwyler:1984} leads to results that contradict the power
counting rules. Subsequently the heavy-baryon formulation of the chiral Lagrangian
was suggested by Jenkins and Manohar \cite{Jenkins:Manohar:1991}. Whereas the chiral power
counting rules are realized transparently, manifest Lorentz invariance is given up in that scheme.
Computations for the baryon octet masses \cite{Bernard:Kaiser:Meissner:1993,Borasoy:Meissner:1997} do
not appear to be
convergent in the heavy-baryon formulation once the strange quark sector is included. The
convergence was improved by introducing a finite cutoff into the
loop functions \cite{Donoghue:Holstein:1998,Donoghue:Holstein:Borasoy:1999,Borasoy:Holstein:Lewis:Ouimet:2002}.
Clearly, alternative schemes are desirable.

This work aims at introducing a partial summation scheme, the construction of which is
guided by covariance and analyticity. Various manifestly Lorentz invariant formulations
of chiral perturbation theory were
suggested that recover the power counting rules \cite{Ellis:Tang:1998,Becher:Leutwyler:1999,Lutz:2000,Gegelia:Japaridze:1999,Fuchs:Gegelia:Japaridze:Scherer:2003,Lutz:Kolomeitsev:2002,Hacker:Wies:Gegelia:Scherer:2005,Bernard:Hemmert:Meissner:2005}.
All such schemes are bound to reproduce computations performed within the heavy-baryon formalism.
The motivation for the search for alternatives stems in part from the quest of summation
schemes that enjoy improved convergence properties. Some of the proposed schemes have been applied
to the evaluation of the baryon octet masses at the one-loop order. The infrared
scheme (IR) introduced by Becher and Leutwyler \cite{Becher:Leutwyler:1999} was used by
Ellis and Torikoshi \cite{Ellis:Torikoshi:1999}, however, finding no convincing convergence properties.
Similarly the extended on-mass shell scheme (EOMS)
introduced by Gegelia and Japaridze  \cite{Gegelia:Japaridze:1999} suffers from unacceptably large
subleading order terms \cite{Lehnhart:Gegelia:Scherer:2005}.

It is the purpose of the present work
to perform computations based on the scheme proposed in \cite{Lutz:2000,Lutz:Kolomeitsev:2002}.
We will evaluate the baryon octet and decuplet self energies at the one-loop level and study
the convergence properties of the minimal chiral subtraction scheme
($\chi$-MS) \cite{Lutz:2000,Lutz:Kolomeitsev:2002}. The latter is
based on the Passarino-Veltman reduction \cite{Passarino:Veltman:1979} supplemented by a
minimal subtraction scheme. It suggests a natural partial summation of the chiral expansion.

It is proven in the Appendix of the present work
that given any one-loop integral that arises when computing one-baryon processes it is
sufficient to renormalize the scalar master-loop functions of the
Passarino-Veltman reduction in a manner that the latter are compatible with the expectation
of chiral counting rules.
Within the $\chi$-MS scheme the empirical octet and decuplet masses can be reproduced accurately
with a small residual dependence on an
infrared renormalization scale only. Good convergence properties are found for natural
values of the infrared scale. A prediction for the strange-quark matrix element of
the nucleon is made.

\section{Relevant chiral interaction terms}

We collect the terms of the chiral Lagrangian that determine the leading orders of
baryon octet and decuplet self energies \cite{Krause:1990,Bernard:Hemmert:Meissner:2003}.
Up to chiral order $Q^2$ the baryon propagators follow from
\begin{eqnarray}
{\mathcal L} &=&  \mathrm{tr} \,\Big( \bar{B}\,\big[i\, \slash{\partial}\,-
\stackrel{\circ}{M}_{[8]}\big]\,B \Big)
\nonumber \\
&-& \mathrm{tr}\,\Big(\bar{\Delta}_\mu \cdot \Big(\big[i\,\slash{\partial}\,
-\stackrel{\circ}{M}_{[10]}\big]\,g^{\mu\nu} -i\,(\gamma^\mu \partial^\nu + \gamma^\nu \partial^\mu)
+ \gamma^\mu\,\big[i\,\slash{\partial} + \stackrel{\circ}{M}_{[10]}\big]\,\gamma^\nu \Big)\,\Delta_\nu\Big)
\nonumber \\
& -& 2\,d_0\, \mathrm{tr}\Big(\bar{\Delta}_\mu \cdot \Delta^\mu \Big)\, \mathrm{tr}\Big(\chi_0\Big)
-2\,d_D\, \mathrm{tr}\Big( (\bar{\Delta}_\mu \cdot \Delta^\mu)\, \chi_0\Big)
\nonumber \\
&+&2\,b_0 \,\mathrm{tr}\Big(\bar{B}\,B\Big)\, \mathrm{tr}\Big(\chi_0\Big)
+ 2\,b_F\,\mathrm{tr}\Big(\bar{B}\,[\chi_0,B]\Big) +
2\,b_D\,\mathrm{tr}\Big(\bar{B}\,\{\chi_0,B\}\Big) \,,
\nonumber\\
\nonumber\\
&& \chi_0 = \left( \begin{array}{ccc}
m_\pi^2 & 0 & 0 \\
0 & m_\pi^2 & 0 \\
0 & 0 & 2\,m_K^2-m_\pi^2
\end{array}\right)\,.
\label{chiral-L}
\end{eqnarray}
We  assume perfect isospin symmetry through out
this work. The  fields are decomposed
into isospin multiplets
\begin{eqnarray}
\Phi &=& \tau \cdot  \pi + \alpha^\dagger \!\cdot \! K + K^\dagger
\cdot \alpha  + \eta \,\lambda_8 \;,
\nonumber\\
\sqrt{2}\,B &=&  \alpha^\dagger \!\cdot \! N+\lambda_8 \,\Lambda+ \tau \cdot \Sigma
 +\Xi^t\,i\,\sigma_2 \!\cdot \!\alpha   \, ,
\end{eqnarray}
with the  Gell-Mann matrices, $\lambda_i$, and the isospin doublet fields
$K =(K^+,K^0)^t $ and $\Xi = (\Xi^0,\Xi^-)^t$.
The isospin Pauli matrices $\sigma=(\sigma_1,\sigma_2,\sigma_3)$ act
exclusively in the space of isospin doublet fields $(K,N,\Xi)$ and
the matrix valued isospin doublet $\alpha$,
\begin{eqnarray}
&& \alpha^\dagger =
 {\textstyle{1\over\sqrt{2}}}\left( \lambda_4+i\,\lambda_5 ,
\lambda_6+i\,\lambda_7 \right) \;,\;\;\;\tau =
(\lambda_1,\lambda_2,\lambda_3)\;.
 \label{def-alpha}
\end{eqnarray}

The tree-level expression for the baryon mass shifts are recalled. For the octet and decuplet
states (\ref{chiral-L}) implies
\begin{eqnarray}
&& \Delta M^{(2)}_{N}=  -2\,m_\pi^2\,(b_0+2\,b_F) - 4\,m_K^2\, (b_0+b_D-b_F)\,,
\nonumber\\
&& \Delta M^{(2)}_{\Sigma} - \Delta M^{(2)}_{\Lambda}=+{\textstyle{16\over 3}}\, b_D\, (m_K^2-m_\pi^2)\,,
\nonumber\\
&& \Delta M^{(2)}_{\Xi} - \Delta M^{(2)}_{N}=- 8\, b_F \,(m_K^2-m_\pi^2)\,,
\nonumber\\
&& \Delta  M^{(2)}_\Xi- \Delta M^{(2)}_\Sigma = - 4\,(b_D+b_F)\,(m_K^2-m_\pi^2)
\,,
\label{Q2-octet}
\end{eqnarray}
and
\begin{eqnarray}
&& \Delta M^{(2)}_{\Delta}
= - 2\,(d_0+d_D)\,m_\pi^2-4\, d_0\,m_K^2\,,
\nonumber\\
&& \Delta M^{(2)}_\Sigma- \Delta M^{(2)}_\Delta
= - {\textstyle{4\over 3}}\,d_D \,(m_K^2-m_\pi^2) \,,
\nonumber \\
&& \Delta M^{(2)}_{\Xi}- \Delta M^{(2)}_\Sigma
= -  {\textstyle{4\over 3}}\,d_D\,(m_K^2-m_\pi^2) \,,
\nonumber \\
&& \Delta M^{(2)}_{\Omega}-\Delta M^{(2)}_\Xi
= -  {\textstyle{4\over 3}}\,d_D\,(m_K^2-m_\pi^2) \,.
\label{Q2-decuplet}
\end{eqnarray}
At tree level the parameters $b_D, b_F$ and $d_D$ can be determined by the mass differences of the
baryon states:
\begin{eqnarray}
b_D \simeq + 0.06 \,{\rm GeV}^{-1} \,, \qquad b_F \simeq - 0.21 \,{\rm GeV}^{-1} \,, \qquad
d_D \simeq - 0.48 \,{\rm GeV}^{-1} \,.
\label{tree-level-parameter}
\end{eqnarray}
It is an amazing result of the tree-level chiral analysis that it yields parameters (\ref{tree-level-parameter})
that are quite consistent with expectations from the large-$N_c$ operator analysis \cite{Dashen,Jenkins}.
At leading order there are three independent parameters, $b_0,b_D$ and $b_F$. It holds
\begin{eqnarray}
b_D+b_F = {\textstyle {1\over 3}}\,d_D \,, \qquad d_0 = b_0\,.
\label{large-Nc-b}
\end{eqnarray}

The evaluation of the baryon self energies to order $Q^3$ probes the meson-baryon vertices
\begin{eqnarray}
{\mathcal L} &=& \frac{F}{2f}\, \mathrm{tr} \,\Big( \bar{B}\, \gamma_5 \gamma^\mu \,[\partial_\mu \Phi,\,B] \Big)
+ \frac{D}{2f}\, \mathrm{Tr}\,\Big( \bar{B}\, \gamma_5 \gamma^\mu \{\partial_\mu \Phi,B\} \Big)
\nonumber\\
&-& \frac{C}{2f}\, \mathrm{tr}\,\Big(\bar{\Delta}_\mu \cdot (\partial_\nu \Phi)\,
\big[g^{\mu\nu}-\frac{1}{2}\,Z\,\gamma^\mu \,\gamma^\nu\big]\, B + \mathrm{h.c.} \Big)
\nonumber\\
&-& \frac{H}{2f} \mathrm{tr} \,\Big(\big[ \bar{\Delta}^\mu \cdot  \gamma_5\,\gamma_\nu\,
\Delta_\mu \big]\,(\partial^\nu \Phi)\,\Big)\,,
\label{chiral-FD}
\end{eqnarray}
where we apply the notations of \cite{Lutz:Kolomeitsev:2002}. We use $f = 92.4$ MeV  in this work.
The values of the coupling constants $F,D,C$ and $H$ may be correlated by a large-$N_c$ operator
analysis \cite{Dashen,Jenkins:Manohar}. At leading order the coupling constants can be expressed in
terms of $F$ and $D$ only. We employ the values for $F$ and $D$ as suggested in \cite{Okun,Lutz:Kolomeitsev:2002}.
All together we use
\begin{eqnarray}
&& F = 0.45 \,, \qquad D= 0.80 \,, \qquad
 H= 9\,F-3\,D \,,\qquad C=2\,D \,,
\label{large-Nc}
\end{eqnarray}
in this work. We take the parameter $Z=0.72$ from a detailed coupled-channel study of meson-baryon
scattering that was based on the chiral Lagrangian \cite{Lutz:Kolomeitsev:2002}.

\section{Baryon octet self energies}

It is straightforward to evaluate the one-loop fluctuation of the baryon octet states as implied
by the interaction vertices specified in (\ref{chiral-FD}). There are two types of contributions that
are characterized by intermediate states involving the octet [8] or decuplet [10]
baryons. For an arbitrary dimension $d$ we write:
\begin{eqnarray}
&&\Sigma^{\rm{loop}}_{B \in [8]}(p)= \sum_{Q\in [8]}
 \int \frac{d^d k}{(2\pi)^d} \,\frac{i\,\mu^{4-d}_{\,UV}\,}{k^2-m_Q^2+i\,\epsilon }\,\Bigg[
\sum_{R\in [8]}\left(\frac{G^{(B)}_{QR}}{2\,f}\right)^2 \gamma_5 \slash{k} \,S_R(p-k)\,
\gamma_5 \slash{k}
\nonumber\\
&& \qquad \qquad \qquad
+ \sum_{R\in [10]}\left(\frac{G^{(B)}_{QR}}{2\,f}\right)^2
\gamma_0\,\Gamma^\dagger_\mu(k)\,\gamma_0 \,S^{\mu \nu}_R(p-k)\, \Gamma^\nu(k) \Bigg]\,,
\nonumber\\
\nonumber\\
&& S^{\mu\nu}_R(p)=\frac{-1}{\slash{p}-M_R + i\epsilon}
\left(g^{\mu\nu} -\frac{\gamma^\mu \,\gamma^\nu}{d-1}-\frac{(d-2)\,p^\mu p^\nu}{(d-1)M_R^2}
+\frac{p^\mu\gamma^\nu -p^\nu \gamma^\mu}{(d-1)\, M_R}\right)\,,
\nonumber\\
&& S_R(p) = \frac{1}{\slash{p}-M_R+i\,\epsilon} \,, \qquad \qquad
\Gamma_\mu(k) =k_\mu-\frac{Z}{2}\,\gamma_\mu\,\slash{k}\,,
\label{octet-self}
\end{eqnarray}
with the notation for the meson-baryon coupling constants $G^{(B)}_{Q R}$ suggested
in  \cite{Lutz:Kolomeitsev:2002}. We assume perfect isospin symmetry in this work. All
coupling constants required in (\ref{octet-self}) are recalled in Tab. \ref{tab:octet-couling} where
we apply the phase convention for the isospin states given in
\cite{Lutz:Kolomeitsev:2002,Kolomeitsev:Lutz:2004}. The meson and baryon masses $m_Q$ and $M_R$
in the propagators are
assumed to be physical, i.e. a partial summation is assumed for the propagators.
The parameter $\mu_{\,UV}$ is the ultraviolet scale of dimensional regularization.

\begin{table}[t]
\rescale
\setlength{\tabcolsep}{1.4mm}
\setlength{\arraycolsep}{5.0mm}
\renewcommand{\arraystretch}{1.2}
\begin{center}
\begin{tabular}{|l|l|l|l|}
\hline
$G^{(N)}_{\pi N}=\sqrt{3}\,(D+F) $&
$G^{(\Lambda)}_{\pi \Sigma}=2\,D $ &
$G^{(\Sigma)}_{\pi \Lambda}=\frac{2\,D}{\sqrt{3}} $&
$G^{(\Xi)}_{\pi \Xi}=-\sqrt{3}\,(D-F) $ \\
%%%%%%%%%%%%%%%%%%%%%%%%%%%%%%%%%%
$G^{(N)}_{\eta N}=-\frac{D-3\,F}{\sqrt{3}} $&
$G^{(\Lambda)}_{\bar{K} N}=-\sqrt{\frac{2}{3}}(D+3F)$ &
$G^{(\Sigma)}_{\pi \Sigma}=-\sqrt{8}\,F $&
$G^{(\Xi)}_{\bar{K} \Lambda}=-\frac{D-3\,F}{\sqrt{3}} $\\
%%%%%%%%%%%%%%%%%%%%%%%%%%%%%%%%%%
$G^{(N)}_{K \Lambda}=-\frac{D+3\,F}{\sqrt{3}} $&
$G^{(\Lambda)}_{\eta \Lambda}=-\frac{2\,D}{\sqrt{3}}$ &
$G^{(\Sigma)}_{\bar{K} N}=\sqrt{2}\,(D-F) $&
$G^{(\Xi)}_{\bar{K} \Sigma}=-\sqrt{3}\,(D+F) $\\
%%%%%%%%%%%%%%%%%%%%%%%%%%%%%%%%%%
$G^{(N)}_{K \Sigma}=\sqrt{3}\,(D-F) $&
$G^{(\Lambda)}_{K \Xi}=\sqrt{\frac{2}{3}}(D-3\,F) $ &
$G^{(\Sigma)}_{\eta\Sigma}=\frac{2\,D}{\sqrt{3}} $&
$G^{(\Xi)}_{\eta \Xi}=-\frac{D+3\,F}{\sqrt{3}} $\\
%%%%%%%%%%%%%%%%%%%%%%%%%%%%%%%%%%
& & $G^{(\Sigma)}_{K \Xi}=\sqrt{2}(D+F) $&  \\ \hline
%%%%%%%%%%%%%%%%%%%%%%%%%%%%%%%%%%%%%%%%%%%%%%%%%%
%%%%%%%%%%%%%%%%%%%%%%%%%%%%%%%%%%%%%%%%%%%%%%%%%%
$G^{(N)}_{\pi \Delta}=2\,C $&
$G^{(\Lambda)}_{\pi \Sigma}=-\sqrt{3}\,C $ &
$G^{(\Sigma)}_{\pi \Sigma}=-\sqrt{\frac{2}{3}}\,C $&
$G^{(\Xi)}_{\pi \Xi}=-C $ \\
%%%%%%%%%%%%%%%%%%%%%%%%%%%%%%%%%%
$G^{(N)}_{K \Sigma}=C $&
$G^{(\Lambda)}_{K \Xi}=-\sqrt{2}\,C$ &
$G^{(\Sigma)}_{\bar K \Delta}=-\sqrt{\frac{8}{3}}\,C $&
$G^{(\Xi)}_{\eta \Xi}=-C $\\
%%%%%%%%%%%%%%%%%%%%%%%%%%%%%%%%%%
&
&
$G^{(\Sigma)}_{\eta\Sigma}=C $&
$G^{(\Xi)}_{\bar{K} \Sigma}=C$\\
%%%%%%%%%%%%%%%%%%%%%%%%%%%%%%%%%%
&
&
$G^{(\Sigma)}_{K \Xi}=-\sqrt{\frac{2}{3}}\,C $&
$G^{(\Xi)}_{K \Omega}=-\sqrt{2}\,C $  \\ \hline
\end{tabular}
\end{center}
\caption{Meson-baryon coupling constants $G^{(B)}_{QR}$ with $B\in [8] $ defined with respect to
isospin states \cite{Lutz:Kolomeitsev:2002}. The upper blocks specify the coupling constants
for $R\in [8]$, the lower blocks the ones for $R\in [10]$.}
\label{tab:octet-couling}
\end{table}

It is long known that the expression (\ref{octet-self}) as it stands
is at odds with chiral power counting rules \cite{Gasser:Sainio:Svarc:1988}. Close to the
baryon mass the one-loop expression  should carry minimal chiral order $Q^3$.  The application of
dimensional regularization in combination with the MS  renormalization scheme
leads to contributions of order $Q^0$ and $Q^2$. Any manifest
Lorentz invariant formulation of chiral perturbation theory takes (\ref{octet-self}) as the
starting point of the renormalization program
\cite{Becher:Leutwyler:1999,Lutz:2000,Gegelia:Japaridze:1999}. Therefore it us useful to
simplify first the expression (\ref{octet-self}). Applying the
Passarino-Veltman reduction \cite{Passarino:Veltman:1979} we obtain the following form
\begin{eqnarray}
&&\Sigma_{B\in [8]}^{\rm{loop}}(p)= \sum_{Q\in [8], R\in [8]}
\left(\frac{G_{QR}^{(B)}}{2\,f} \right)^2 \Big[a^{[8]}_{QR}(p)\,I_R
+b^{[8]}_{QR}(p)\,I_Q + c^{[8]}_{QR}(p)\,I_{QR}(p^2)\Big]
\nonumber\\
&& \qquad \!\! +\sum_{Q\in [8], R\in [10]}
\left(\frac{G_{QR}^{(B)}}{2\,f} \right)^2 \Big[a^{[10]}_{QR}(p)\,I_R
+b^{[10]}_{QR}(p)\,I_Q + c^{[10]}_{QR}(p)\,I_{QR}(p^2)\Big] \,,
\label{PV-octet}
\end{eqnarray}
in terms of the invariant master loop functions
\begin{eqnarray}
&&I_Q = \int \frac{d^d k}{(2\pi)^d} \,\frac{i\,\mu^{4-d}_{\,UV}\,}{k^2-m_Q^2+i\,\epsilon } \,,\qquad
I_R = \int \frac{d^d k}{(2\pi)^d} \,\frac{i\,\mu^{4-d}_{\,UV}\,}{k^2-M_R^2+i\,\epsilon } \,,
\nonumber\\
&&I_{QR}(p^2) =
\int \frac{d^d k}{(2\pi)^d} \,\frac{-i\,\mu^{4-d}_{\,UV}\,}{k^2-m_Q^2+i\,\epsilon }
\frac{1}{(p-k)^2-M_R^2+i\,\epsilon}\,.
\label{def-master-loop}
\end{eqnarray}
The coefficient functions are readily derived. The baryon-octet intermediate states define
\begin{eqnarray}
&&a_{QR}^{[8]}(p) =-M_R -\frac{M_R^2+p^2}{2\,p^2}\,\slash{p}\,, \qquad \quad
b_{QR}^{[8]}(p) =\frac{M_R^2-p^2}{2\,p^2}\,\slash{p}\,,
\nonumber\\
&&c_{QR}^{[8]}(p) = m_Q^2\,M_R -\frac{(M_R^2-p^2)^2-m_Q^2\,(M_R^2+p^2)}{2\,p^2}\,\slash{p} \,.
\label{octet-result-8}
\end{eqnarray}
The baryon-decuplet intermediate states lead to
\begin{eqnarray}
&&a_{QR}^{[10]}(p) =\frac{d-2}{8\,(d-1)\,M^2_R\,p^2}
\Big[ 2\,M_R\,p^2\,\Big(m_Q^2-M_R^2-p^2\Big)
\nonumber\\
&&\quad +\Big\{-m_Q^4+2\,(M_R^2+p^2)\,m_Q^2-(M_R^2-p^2)^2-\frac{4}{d}\, M_R^2\,p^2\Big\}\,\slash{p}
\Big]
\nonumber\\
&&b_{QR}^{[10]}(p) =\frac{d-2}{8\,(d-1)\,M^2_R\,p^2}
\Big[ 2\,M_R\,p^2\,\Big( \frac{4\,Z-d\,(Z^2-3)-6}{d-2}\,m_Q^2+M_R^2-p^2\Big)
\nonumber\\
&& \quad +\Big\{ m_Q^4
+\Big(\Big(\frac{4\,(Z-1)^2}{d}-2\,(Z^2-2)\Big)\,p^2-2\,M_R^2 \Big)\,m_Q^2
+M_R^4-(p^2)^2\Big\}\,\slash{p} \Big]
\nonumber\\
&&c_{QR}^{[10]}(p) =\frac{d-2}{8\,(d-1)\,M^2_R\,p^2}\,
\Big[ 2\,M_R\,p^2\,\Big(-m_Q^4+2\,(M_R^2+p^2)\,m_Q^2
\nonumber\\
&&\quad -(M_R^2-p^2)^2\Big)
+\Big\{m_Q^6-3\,(M_R^2+p^2)\,m_Q^4
\nonumber\\
&&\quad +\Big(3\,M_R^4+2\,p^2\,M_R^2+3\,(p^2)^2 \Big)\,m_Q^2
 -(M_R^2-p^2)^2\,(M_R^2+p^2)
\Big\} \,\slash{p} \Big] \,.
\label{octet-result-10}
\end{eqnarray}

\section{Baryon decuplet self energies}

We turn to the one-loop fluctuations of the baryon decuplet states. There are two terms
induced by intermediate baryon octet and decuplet states. We write
\begin{eqnarray}
&&\!\!\Sigma_{B\in [10]}^{\mu \nu,\,\rm{loop}}(p)= \!\sum_{Q\in [8]}
 \int \!\frac{d^d k}{(2\pi)^d} \,\frac{-i\,\mu^{4-d}_{\,UV}\,}{k^2-m_Q^2+i\,\epsilon }\,\Bigg[
 \sum_{R\in [10]}\!\!\left( \frac{G^{(B)}_{QR}}{2\,f}\right)^2
\!\!\!\gamma_5\slash{k} \,S^{\mu \nu}_R(p-k)\, \gamma_5\slash{k}
\nonumber\\
&& \qquad \qquad \qquad
+ \sum_{R\in [8]}\left(\frac{G^{(B)}_{QR}}{2\,f}\right)^2 \!\!\Gamma^\mu(k) \,S_R(p-k)\,
\gamma_0\,\Gamma^{\nu,\dagger}(k) \,\gamma_0 \Bigg]\,,
\label{decuplet-self}
\end{eqnarray}
where the building blocks of (\ref{decuplet-self}) are specified in (\ref{octet-self}).
A list with coupling constants, $G^{(B)}_{QR}$, required in (\ref{decuplet-self}) is provided
in Tab. \ref{tab:decuplet-coupling}. Perfect isospin symmetry is assumed.

\begin{table}[t]
\rescale
\setlength{\tabcolsep}{1.4mm}
\setlength{\arraycolsep}{5.0mm}
\renewcommand{\arraystretch}{1.2}
\begin{center}
\begin{tabular}{|l|l|l|l|}
\hline
$G^{(\Delta)}_{\pi N}=\sqrt{2}\,C $&
$G^{(\Sigma)}_{\pi \Lambda}=-C $&
$G^{(\Xi)}_{\pi \Xi}=-C $ &
$G^{(\Omega)}_{\bar K \Xi}=-2\,C $ \\
%%%%%%%%%%%%%%%%%%%%%%%%%%%%%%%%%%
$G^{(\Delta)}_{K \Sigma}=-\sqrt{2}\,C  $&
$G^{(\Sigma)}_{\pi \Sigma}=-\sqrt{\frac{2}{3}}\,C $&
$G^{(\Xi)}_{\bar{K} \Lambda}=C $ &\\
%%%%%%%%%%%%%%%%%%%%%%%%%%%%%%%%%%
&
$G^{(\Sigma)}_{\bar{K} N}=\sqrt{\frac{2}{3}}\,C $&
$G^{(\Xi)}_{\bar{K} \Sigma}=C $ &\\
%%%%%%%%%%%%%%%%%%%%%%%%%%%%%%%%%%
&
$G^{(\Sigma)}_{\eta\Sigma}=C $&
$G^{(\Xi)}_{\eta \Xi}=-C $&\\
%%%%%%%%%%%%%%%%%%%%%%%%%%%%%%%%%%
&
$G^{(\Sigma)}_{K \Xi}=-\sqrt{\frac{2}{3}}\,C $
&
&  \\ \hline
%%%%%%%%%%%%%%%%%%%%%%%%%%%%%%%%%%%%%%%%%%%%%%%%%%
%%%%%%%%%%%%%%%%%%%%%%%%%%%%%%%%%%%%%%%%%%%%%%%%%%
$G^{(\Delta)}_{\pi \Delta}=-\sqrt{\frac{5}{3}}\,H $&
$G^{(\Sigma)}_{\pi \Sigma}=\frac{\sqrt{8}}{3}\,H $&
$G^{(\Xi)}_{\pi \Xi}=-\sqrt{\frac{1}{3}}\,H $&
$G^{(\Omega)}_{\bar K \Xi}=-\frac{2}{\sqrt{3}}\,H $ \\
%%%%%%%%%%%%%%%%%%%%%%%%%%%%%%%%%%
$G^{(\Delta)}_{\eta \Delta}=-\sqrt{\frac{1}{3}}\,H $&
$G^{(\Sigma)}_{\bar K \Delta}=-\frac{\sqrt{8}}{3}\,H $&
$G^{(\Xi)}_{\bar K \Sigma}=-\frac{2}{\sqrt{3}}\,H $&
$G^{(\Omega)}_{\eta \Omega}=\frac{2}{\sqrt{3}}\,H $\\
%%%%%%%%%%%%%%%%%%%%%%%%%%%%%%%%%%
$G^{(\Delta)}_{K \Sigma}=-\sqrt{\frac{2}{3}}\,H $&
$G^{(\Sigma)}_{\eta \Sigma}=0 $&
$G^{(\Xi)}_{K \Omega}=-\sqrt{\frac{2}{3}}\,H $ &
\\
%%%%%%%%%%%%%%%%%%%%%%%%%%%%%%%%%%
&
$G^{(\Sigma)}_{K \Xi}=\frac{\sqrt{8}}{3}\,H $  &
$G^{(\Xi)}_{\eta \Xi}=\frac{1}{\sqrt{3}}\,H $  &
\\ \hline
\end{tabular}
\end{center}
\caption{Meson-baryon coupling constants $G^{(B)}_{QR}$ with $B\in [10] $ defined with respect to
isospin states \cite{Lutz:Kolomeitsev:2002,Kolomeitsev:Lutz:2004}. The upper blocks specify the coupling constants
for $R\in [8]$, the lower blocks the ones for $R\in [10]$.}
\label{tab:decuplet-coupling}
\end{table}

The self energy
tensor, $\Sigma_{\mu \nu}(p)$, determines the dressed propagator, $S_{\mu \nu}(p)$,
by means of the Dyson equation
\begin{eqnarray}
S_{\mu \nu} (p) = S^{(0)}_{\mu \nu}(p) + S^{(0)}_{\mu \alpha}(p)\,\Sigma^{\alpha \beta}(p)\,S_{\beta \nu}(p)\,,
\end{eqnarray}
where the bare propagator, $S_{\mu \nu}^{(0)}(p)$, follows from the expression in (\ref{octet-self})
upon using the bare decuplet mass.
The Dirac-Lorentz structure of a spin three-half particle causes a little complication.
It is convenient to decompose the self energy into a complete set of tensors \cite{Lutz:Korpa:2002}
defined for arbitrary dimension $d$:
\begin{eqnarray}
&& P^{\mu \nu}_{\frac{3}{2}\pm} (p) = \Big(g^{\mu \nu}-\frac{p^\mu\,p^\nu}{p^2} \Big)\,P_\pm(p)
-V^\mu(p)\,P_\mp(p) \,V^\nu (p)\,,
\nonumber\\
&& P^{\mu \nu}_{\frac{1}{2}\pm,11} = V^\mu(p)\,P_\mp(p)\,V^\nu (p)\,, \qquad
P^{\mu \nu}_{\frac{1}{2}\pm,12} = V^\mu(p)\,P_\mp(p)\,\frac{p^\nu}{\sqrt{p^2}}\,,
\nonumber\\
&&P^{\mu \nu}_{\frac{1}{2}\pm,21} = \frac{p^\mu}{\sqrt{p^2}}\,P_\mp(p)\,V^\nu (p)\,,
\qquad \;\,\,
P^{\mu \nu}_{\frac{1}{2}\pm,22} = \frac{p^\mu}{\sqrt{p^2}}\,P_\mp(p)\,\frac{p^\nu}{\sqrt{p^2}}\,,
\nonumber\\
&& P_\pm (p) = \frac{1}{2}\,\Big( 1\pm \frac{\slash{p}}{\sqrt{p^2}}\Big) \,, \qquad \qquad \;\;
V^\mu(p)= \frac{1}{\sqrt{d-1}}\,\Big( \gamma^\mu- \frac{\slash{p}\,p^\mu}{p^2}\Big)\,.
\label{def-tensors}
\end{eqnarray}
Any Dirac-Lorentz tensor, $A_{\mu \nu}(p)$, that depends on a single 4-momentum only
can be represented as follows
\begin{eqnarray}
&&A^{\mu \nu}(p) = \sum_{\pm } \,A^{\frac{3}{2}\pm}(p^2)\,P_{\frac{3}{2}\pm}^{\mu \nu}(p) +
\sum_{ij,\pm } \,A^{\frac{1}{2}\pm}_{ij}(p^2)\,P_{\frac{1}{2}\pm,ij}^{\mu \nu}(p) \,,
\label{tensor-expand}\\
&& A^{\frac{3}{2}\pm}(p^2) =  \frac{2}{d\,(d-2)}\,
{\tr} P_{\frac{3}{2}\pm}^{\mu \nu}(p)\,A_{\nu \mu}(p)
\,, \quad
 A^{\frac{1}{2}\pm}_{ij}(p^2) =  \frac{2}{d}\,{\tr} P_{\frac{1}{2}\pm,ji}^{\mu \nu}(p)\,
A_{\nu \mu}(p)\,.
\nonumber
\end{eqnarray}
The information on the decuplet masses is encoded in the spin three-half components of the
self energy. Owing to the projector properties of the tensors introduced in (\ref{def-tensors})
those components are determined by the corresponding components of the self energy tensor. It holds
\begin{eqnarray}
&&S_B^{\frac{3}{2}\pm}(p^2) = -\Big[\sqrt{p^2}\,\mp\stackrel{\circ}{M}_{[10]}
- \Sigma_B^{\frac{3}{2}\pm}(p^2) \Big]^{-1} \,,
\nonumber\\
&& M_B = \stackrel{\circ}{M}_{[10]} +\,\Re \Sigma_B^{\frac{3}{2}+}(M_B^2) \,,
\end{eqnarray}
where we apply the quasi-particle definition of the decuplet masses.

Like for the baryon-octet self energies it is useful to derive simplified and explicit
representations of the spin three-half components of the decuplet self energies.
Applying the Passarino-Veltman reduction we seek a representation for the one-loop
contribution to the decuplet self energy of the form
\begin{eqnarray}
&&\Sigma_{B\in [10]}^{\rm{loop}}(p)= \sum_{Q\in [8], R\in [8]}
\left(\frac{G_{QR}^{(B)}}{2\,f} \right)^2 \Big[a^{[8]}_{QR}(p)\,I_R
+b^{[8]}_{QR}(p)\,I_Q + c^{[8]}_{QR}(p)\,I_{QR}(p^2)\Big]
\nonumber\\
&& \qquad +\sum_{Q\in [8], R\in [10]}
\left(\frac{G_{QR}^{(B)}}{2\,f} \right)^2 \Big[a^{[10]}_{QR}(p)\,I_R
+b^{[10]}_{QR}(p)\,I_Q + c^{[10]}_{QR}(p)\,I_{QR}(p^2)\Big] \,,
\label{PV-decuplet}
\end{eqnarray}
where the master loop functions were already introduced in (\ref{def-master-loop}).
For notational convenience we suppress the index $\frac{3}{2}+$ in the self energies.
It is straightforward to derive the various components
\begin{eqnarray}
&&a^{[8]}_{QR}(p) = \frac{M_R}{8\,(d-1)\,p^2\,p^2} \,\Big[ 2\,p^2\,
\Big( m_Q^2-M_R^2-p^2\Big)
\nonumber\\
&& \quad+\Big\{-m_Q^4 + 2\,(M_R^2 + p^2)\, m_Q^2 -(M_R^2-p^2)^2
- \frac{4\,M_R^2\, p^2}{d} \Big\}\,\frac{\sqrt{p^2}}{M_R}\,\Big] \,,
\nonumber\\
&&b^{[8]}_{QR}(p) = \frac{M_R}{8\,(d-1)\,p^2\,p^2} \,\Big[ 2\,p^2\,
\Big( -m_Q^2+M_R^2-p^2\Big)
\nonumber\\
&& \quad+\Big\{m_Q^4 - 2\,\Big(M_R^2 + 2\,p^2 - \frac{2\,p^2}{d}\Big)\,m_Q^2
+ M_R^4-p^2\,p^2 \Big\}\,\frac{\sqrt{p^2}}{M_R}\,\Big] \,,
\nonumber\\
&&c^{[8]}_{QR}(p) = \frac{M_R}{8\,(d-1)\,p^2\,p^2} \,\Big[ 2\,p^2\,
\Big( -m_Q^4 + 2\,(M_R^2+p^2)\, m_Q^2 - (M_R^2-p^2)^2 \Big)
\nonumber\\
&& \quad+\Big\{m_Q^6 - 3\,(M_R^2 + p^2 )\, m_Q^4
+ (3\,M_R^4+2\,M_R^2\, p^2 + 3\,p^2\,p^2)\, m_Q^2   \nonumber\\
&& \quad
- (M_R^2-p^2)^2 \,(M_R^2+p^2) \Big\}\,\frac{\sqrt{p^2}}{M_R}\,\Big] \,,
\label{decuplet-result-8}
\end{eqnarray}
and

\begin{eqnarray}
&&a^{[10]}_{QR}(p) = \frac{1}{8\,(d-1)^2\,M_R\,p^2\,p^2} \,\Big[ 2\,p^2\,
\Big(  (d-4)\, (M_R^2+p^2)\, m_Q^2-(d-2)\, m_Q^4
\nonumber\\
&&\quad +  4\,d\,(3-2\,d)\, M_R^2\, p^2  + 2\,(M_R^4-6\,p^2\, M_R^2 +p^2\,p^2)
+ \frac{8\,M_R^2\, p^2}{d}\Big)
\nonumber\\
&& \quad+\Big\{  -2\,(d-2)\,(M_R^2+p^2)\, m_Q^4 +
2\,\Big((d-2)\,(M_R^4+p^2\,p^2) - 4\,p^2 \,M_R^2\Big)\, m_Q^2
\nonumber\\
&&\quad  - (M_R^2+p^2) \,\Big(4\,d^2\,M_R^2\,p^2+ d\,(M_R^4-14\,p^2\, M_R^2 +p^2\,p^2 )
\nonumber\\
&& \quad  - 2\,(M_R^4-6\,p^2\,M^2+p^2\,p^2)-\frac{8\,p^2\, M_R^2}{d} \Big)\Big\}\,
\frac{\sqrt{p^2}}{M_R}\,\Big] \,,
\nonumber\\
&&b^{[10]}_{QR}(p) = \frac{1}{8\,(d-1)^2\,M_R\,p^2\,p^2} \,\Big[ 2\,p^2\,
\Big((d-2)\,m_Q^4 + \Big((4-d)\, M_R^2
\nonumber\\
&& \quad +(d-8+\frac{8}{d})\,p^2\Big)\,m_Q^2 -2\,(M_R^4-p^2\,p^2) \Big)
+\Big\{(d-2)\,(M_R^2+p^2 )\, m_Q^4
\nonumber\\
&& \quad  -2\,\Big((d-2)\,M_R^4 + (d-8)\,p^2\, M_R^2 + \frac{4\,p^2\, M_R^2}{d}
- 2\,(d-3 + \frac{2}{d} )\,p^2\,p^2 \Big)\, m_Q^2
\nonumber\\
&& \quad  + (M_R^2-p^2 )\, \Big( (d-2)\,M_R^4 + 2\,d\,(2\,d-5)\,p^2\, M_R^2+ (d-2)\,p^2\,p^2\Big)
\Big\}\,\frac{\sqrt{p^2}}{M_R}\,\Big] \,,
\nonumber\\
&&c^{[10]}_{QR}(p) = \frac{1}{8\,(d-1)^2\,M_R\,p^2\,p^2} \,\Big[ 2\,p^2\,
\Big( (d-2)\, m_Q^6 - 2\,(d-3)\,(M_R^2+p^2)\, m_Q^4
\nonumber\\
&& \quad  + \Big( (d-6)\,(M_R^4+p^2\,p^2)
+2\,d\,(2\,d-5)\,p^2\, M_R^2\Big)\,m_Q^2
\nonumber\\
&& \quad + 2\,(M_R^2-p^2)^2 \,(M_R^2+p^2 )  \Big)
+\Big\{ (d-2)\,(M_R^2+p^2 )\, m_Q^6
\nonumber\\
&& \quad - \Big(3\,(d-2)\,(M_R^4+p^2\,p^2 ) + 2\,(d-6)\,p^2\, M_R^2\Big)\,m_Q^4
\nonumber\\
&& \quad  + (M_R^2+p^2)\,\Big(3\,(d-2)\,(M_R^4+p^2\,p^2) + 2\,d\,(2\,d-7)\,p^2\,
M_R^2\Big)\, m_Q^2
\nonumber\\
&& \quad -(M_R^2-p^2)^2\, \Big( (d-2)\,(M_R^4+p^2\,p^2) + 2\,d\,(2\,d-5)\,p^2\, M_R^2\Big)
\Big\}\,\frac{\sqrt{p^2}}{M_R}\,\Big] \,.
\label{decuplet-result-10}
\end{eqnarray}

\section{Renormalization and power counting}

It is important to  discriminate carefully two
different issues. First, are the chiral Ward identities satisfied and second are the power
counting rules manifest? We point out that the one-loop expressions for the self energies
(\ref{PV-octet}, \ref{PV-decuplet}) are consistent with all chiral Ward identities simply
because whatever symmetry the Lagrangian enjoys dimensional regularization preserves those
at the level of the Green functions. The loop expansion does not cause a violation of the
Ward identities. The task is to devise a renormalization scheme that
preserves the Ward identities but leads at the same time to manifest power counting for the
renormalized loop functions.

We focus on the renormalization for the one-loop expressions $I_Q$,
$I_N$ and $I_{Q R}(p^2)$ introduced in (\ref{def-master-loop}). It is emphasized that those
master loops are the only ultraviolet divergent objects that arise in the computation of
any one-loop diagram if the Passarino-Veltman reduction is applied. Any scalar master loop
function that arises in the Passarino-Veltman reduction that is finite and non-trivial in the
chiral domain behaves as dictated by power counting rules. The latter statement is almost trivial
since for finite integrals dimensional counting is justified as long as performing the loop
integrations commutes with taking the limit of large baryon masses. Scalar integrals that are
trivial in the chiral domain, i.e. those ones that can be Taylor expanded in the soft momenta
may violate the counting rules. This expectation was confirmed by explicit computations
\cite{Lutz:2000}. It is proven in the Appendix of the present work
that given any one-loop integral that arises when computing one-baryon processes it is
sufficient to renormalize the scalar master-loop functions of the
Passarino-Veltman reduction in a manner that the latter are compatible with the expectation
of chiral counting rules.

Thus it is of central importance to consider the  ultraviolet divergent master loop function.
We first recall their well-known properties
for arbitrary space-time dimension $d$. The tadpole loop has the form
\begin{eqnarray}
&& I_R = M_R^2\,\frac{\Gamma (1-d/2)}{(4\pi)^2}
\left(\frac{M_R^2}{4\,\pi \,\mu^2} \right)^{(d-4)/2}
\nonumber\\
&& \quad \;\,= \frac{M_R^2}{(4\,\pi)^2}
\left( -\frac{2}{4-d}+\gamma-1 -\ln (4 \pi)+\ln \left( \frac{M_R^2}{\mu_{UV}^2}\right)
+ {\mathcal O}\left(4-d \right)\right) \;,
\label{n-tadpole}
\end{eqnarray}
where $\gamma $ is the Euler constant.  The expression
for the mesonic tadpole, $I_Q$, follows by replacing the mass $M_R$ in (\ref{n-tadpole})
by the meson mass $m_Q$. In dimensional regularization the divergent part of the
master loop $I_{QR}(p^2)$ is determined unambiguously by the tadpole specified
in (\ref{n-tadpole}). The algebraic identity
\begin{eqnarray}
I_{QR}(0) = \frac{I_R-I_Q}{m_Q^2-M_R^2} \,,
\label{def-algebra-constraint}
\end{eqnarray}
holds for arbitrary values of $d$. If we slightly rewrite the Passarino-Veltman representation,
using the subtracted master loop
\begin{eqnarray}
\Delta I_{QR}(p^2)=I_{QR}(p^2)-I_{QR}(0) \,,
\end{eqnarray}
rather than the original loop function $I_{QR}(p^2)$, the renormalization of the ultraviolet divergencies
is reduced to the consideration of the tadpole terms only. We write
\begin{eqnarray}
&&\Sigma_{B}^{\rm{loop}}(p) \sim  \Delta a_{QR}(p)\,I_R
+\Delta b_{QR}(p)\,I_Q + c_{QR}(p)\,\Delta I_{QR}(p^2) \,,
\nonumber\\
&& \Delta a_{QR}(p) = a_{QR}(p)+ \frac{c_{QR}(p)}{m_Q^2-M_R^2}\,, \quad
\Delta b_{QR}(p) = b_{QR}(p)- \frac{c_{QR}(p)}{m_Q^2-M_R^2} \,.
\label{rewrite-PV}
\end{eqnarray}
It is left to specify the subtracted master loop $\Delta I_{Q R}(p^2)$.
Since it is finite it suffices to recall its form for
$d=4$:
\begin{eqnarray}
\!\!\!\!\Delta I_{Q R}(p^2)&=&\frac{1}{16\,\pi^2}
\left( 1+\left(\frac{1}{2}\,\frac{m_Q^2+M_R^2}{m_Q^2-M_R^2}
-\frac{m_Q^2-M_R^2}{2\,p^2}
\right)
\,\ln \left( \frac{m_Q^2}{M_R^2}\right)
\right.
\nonumber\\
&+&\left.
\frac{p_{Q R}}{\sqrt{p^2}}\,
\left( \ln \left(1-\frac{p^2-2\,p_{Q R}\,\sqrt{p^2}}{m_Q^2+M_R^2} \right)
-\ln \left(1-\frac{p^2+2\,p_{Q R}\sqrt{p^2}}{m_Q^2+M_R^2} \right)\right)
\right)\;,
\nonumber\\
p_{Q R}^2 &=&
\frac{p^2}{4}-\frac{M_R^2+m_Q^2}{2}+\frac{(M_R^2-m_Q^2)^2}{4\,p^2}  \;.
\label{ipin-analytic}
\end{eqnarray}
We emphasize that the master loop function $\Delta I_{QR}(p^2)$ satisfies a
once subtracted dispersion-integral representation
\begin{eqnarray}
&& \Delta I_{Q R}(p^2)= \int_{(m_Q+M_R)^2}^\infty \frac{d s}{\pi}
\,\frac{p^2}{s}\,\frac{\Im \Delta I_{QR}(s)}{s-p^2-i\,\epsilon}\,,
\nonumber\\
&& \Im \Delta I_{QR}(p^2) = \frac{p_{QR}}{8\,\pi \sqrt{s}}\,\Theta \Big(p^2-(M_R+m_Q)^2 \Big)\,,
\label{disp-integral}
\end{eqnarray}
where we still assume $d=4$. Our renormalization scheme will be constructed in a manner
that causal properties like (\ref{disp-integral}) are untouched. This desire motivates the
partial summation we are after in this work.

It is remarked that in the scheme of Becher and
Leutwyler \cite{Becher:Leutwyler:1999} the renormalized expression for the integral
$I_{QR}(p^2)$ does not satisfy a dispersion-integral representation of the form (\ref{disp-integral}).
The corresponding expression reads:
\begin{eqnarray}
&& I_{QR}(p^2) \stackrel{I.R.}{\longrightarrow}  -
\frac{1}{8\pi^2} \frac{\alpha \sqrt{1-\Omega^2}}{1+2\,\alpha\,\Omega + \alpha^2}\,
\arccos \left(-\frac{\alpha+\Omega}{\sqrt{1+2\,\alpha\,\Omega+\alpha^2}}\right)
\nonumber\\
&& \qquad \qquad \quad - \frac{1}{16\,\pi^2}\frac{\alpha\,(\alpha+\Omega)}{1+2\,\alpha\,\Omega+\alpha^2}\,
(2\,\ln\alpha-1)\,,
\nonumber\\
&& \alpha=\frac{m_Q}{M_R}, \hspace{20pt}\quad  \Omega=\frac{p^2-M_R^2-m_Q^2}{2\,M_R\,m_Q} \,,
\label{IR-log-integral}
\end{eqnarray}
which implies that  $\Im I_{QR}(p^2)$ is non-vanishing at $p^2<0$. Such contributions
are unphysical leading to an a-causal dispersion-integral representation. It should be stressed that it is
legitimate to accept this artifact arguing that those structures are far away from the region where
the results are reliable. Nevertheless, it is useful to establish an alternative scheme that does
not violate causality.

We are now well prepared to define our renormalization scheme, which is a slightly
generalized version of the scheme suggested in \cite{Lutz:2000,Lutz:Kolomeitsev:2002}.
The renormalized quantities are introduced with
\begin{eqnarray}
&& I_R= \delta I_R + \bar I_R\,, \quad I_Q = \delta I_Q + \bar I_Q\,, \quad
\Delta I_{QR}(p^2) = \delta I_{QR}+ \bar I_{QR}(p^2)\,.
\label{def-scheme}
\end{eqnarray}
Before specifying the form of the renormalized quantities $\bar I_R$, $\bar I_Q$ and
$\bar I_{QR}(p^2)$ it is useful to discuss possible constraints that may have to be watched.
We argue in fact that it is legitimate to choose the subtraction terms $\delta I_R$, $\delta I_Q$
and $\delta I_{QR}$ almost arbitrarily. This is a consequence of a simple observation: any Ward
identity which one may write down at the one-loop level can be analyzed in the Passarino-Veltman
representation, i.e. the left-hand and right-hand side of the Ward identity are
linear combinations of the master loop functions.
We conclude that the coefficients in front of the master loop functions on the left-hand and
right-hand side must match. This holds under the assumption that
the Passarino-Veltman representation is defined unambiguously.
As a consequence it appears justified to construct $\delta I_R$, $\delta I_Q$
and $\delta I_{QR}$ arbitrarily.

However, a subtle complication arises: the
Passarino-Veltman coefficients may have kinematical singularities. Before renormalization
the latter are superficial since they cancel due to interrelations amongst the master
loop functions at specific kinematical points (see e.g. \cite{Lutz:2000}).
It is advantageous to keep those interrelations as much as possible in order to arrive
at a scheme that suffers from few kinematical singularities only. The occurrence
of kinematical singularities is not an argument against a particular scheme. For instance,
the heavy-baryon formulation is known to suffer severely from those. The presence of
kinematical singularities is a typical phenomena  associated
when restoring chiral power counting rules \cite{Lutz:2000,Lutz:Kolomeitsev:2002}.
It was emphasized in  \cite{Lutz:2000,Lutz:Kolomeitsev:2002} that all
scalar one-loop integrals that are ultraviolet finite confirm the expectation
of naive power counting \cite{Lutz:2000,Lutz:Kolomeitsev:2002}. Thus there is
no need to devise a renormalization as to modify their leading chiral power.
Nevertheless, it still may be advantageous to modify the latter as to eliminate
unwanted kinematical singularities. This is analogous to the heavy-baryon scheme, in
which partial summation can be performed in order to restore the proper analytic structure of
particular contributions. This issue will be discussed further in a forthcoming paper.

We proceed and specify the $\chi$-MS scheme with
\begin{eqnarray}
&& \delta I_Q \;=\frac{m_Q^2}{(4\,\pi)^2}
\left( -\frac{2}{4-d}+\gamma-1 -\ln (4 \pi) \right) \,, \qquad
\nonumber\\
&& \delta I_R \;=\frac{M_R^2}{(4\,\pi)^2}
\left( -\frac{2}{4-d}+\gamma-1 -\ln (4 \pi)+\ln \left( \frac{M_R^2}{\mu_{\,UV}^2}\right) \right)\,,
\nonumber\\
&& \delta I_{QR} = \frac{1}{(4\pi)^2}\,\left(1- 2\,\frac{\mu_{IR}}{M_R}\right)\,.
\label{def-chiMS}
\end{eqnarray}
As a consequence of (\ref{def-chiMS}) the renormalized quantities
$\bar I_R$, $\bar I_Q$ and $\bar I_{QR}(p^2)$ take
the form:
\begin{eqnarray}
&& \bar I_{R} = 0 \,, \qquad \qquad \bar I_Q =\frac{m_Q^2}{(4\,\pi)^2}\,
\ln \left( \frac{m_Q^2}{\mu_{\,UV}^2}\right)    \;,
\nonumber\\
&&  \bar I_{Q R}(M^2_R)=
\frac{1}{8\,\pi^2}\,\frac{\mu_{IR}}{M_R}-\frac{m_Q}{16\,\pi\,M_R}  \left(1-\frac{m_Q^2}{8\,M_R^2} \right)
\nonumber\\
&& \qquad \qquad \,+\,\frac{1}{(4 \pi)^2}\left( 1
-\frac{3}{2}\,\ln \left( \frac{m_Q^2}{M_R^2}\right) \right)
\frac{m_Q^2}{M_R^2}
+{\mathcal O}\left( \frac{m_Q^4}{M_R^4}, d-4 \right) \,.
\label{ren-quantities}
\end{eqnarray}
The renormalized mesonic tadpole $\bar I_Q$ agrees with
the corresponding expression implied by the $\widetilde{MS}$-scheme. The vanishing of the
baryonic tadpole $\bar I_R$ is analogous to what is assumed in the infrared regularization
scheme of Becher and Leutwyler. The crucial element is the presence of an infrared
renormalization scale $\mu_{IR}$ in the expression for $\bar I_{QR}(p^2)$ (see also \cite{KSW}).
It is pointed out that the results (\ref{ren-quantities}) are consistent with
the power counting rules provided that the infrared renormalization scale $\mu_{IR} \sim Q$ is
assigned the chiral power $Q$.

Let us justify the presence of the infrared renormalization scale in $\delta I_{QR}$.
Consider the properties
of the unrenormalized object $\Delta I_{QR}(m_R^2)$ in more detail. Even though it is finite we introduce
a cutoff $\Lambda$ in the
dispersion-integral representation (\ref{disp-integral}) and study the non-relativistic limit with
$\Lambda \ll M_R$:
\begin{eqnarray}
&&\Delta I^{(\Lambda)}_{QR}(M_R^2)=\int^{(\sqrt{m_Q^2+\Lambda^2}+ \sqrt{M_R^2+\Lambda^2}\,)^2}_{(m_Q+M_R)^2}
 \frac{d s}{\pi}\,\frac{M_R^2}{s}\,\frac{\Im I_{QR}(s)}{s-M_R^2}
\nonumber\\
&& \qquad \qquad \;\;\;\;= \frac{1}{4\,\pi^2}\,\int_0^\Lambda
\frac{d l\,l^2}{E^2_Q\,E_R+E^2_R\,E_Q}\,\frac{M_R^2}{(E_Q+E_R)^2-M_R^2}
\nonumber\\
&& \qquad \qquad \;\;\;\;= \frac{1}{8\,\pi^2\,M_R}\,\int_0^\Lambda \frac{d l\,l^2}{m_Q^2+l^2} \left(1+
 {\mathcal O} \left( \frac{ \Lambda^2}{M^2_R}\right) \right)\,,
\nonumber\\
&& E_Q= \sqrt{m_Q^2+l^2}\,, \qquad \qquad E_R= \sqrt{M_R^2+l^2}\,.
\label{cutoff-IQR}
\end{eqnarray}
The result (\ref{cutoff-IQR}) illustrates that the non-relativistic limit with $\Lambda \ll M_R$ introduces an additional
divergence. In order to justify the expansion (\ref{cutoff-IQR}) it is necessary to count $\Lambda \sim Q$. Once
we accept this, power counting rules are manifest. This holds also for cutoff-regularized meson and baryon tadpoles
\cite{Lutz:2000}.
It must be emphasized that the integral $\Delta I_{QR}(p^2)$ is finite a priori, only its non-relativistic
expansion leads to power-like divergencies. As a result the associated renormalization scale must not necessarily
be identified with the renormalization scale characterizing for instance the mesonic tadpole $I_Q$.
We argue that it is legitimate to implement two different cutoff scales $\Lambda_{IR}$ and $\Lambda_{UV}$ with
$\Lambda_{UV} \gg \Lambda_{IR}$. Whereas the ultraviolet cutoff is used to limit meson momenta the second
cutoff is applied to limit baryon three momenta. Though it is possible to work with two different cutoff
parameters it is in practice quite cumbersome to ensure that none of the symmetry constraints is violated.
That is why it is advantageous to rely on dimensional regularization  mimicking the scale scenario.
The ultraviolet cutoff parameter $\Lambda_{UV}$ translates into the ultraviolet renormalization scale
$\mu_{UV}$. The role of the second cutoff  parameter $\Lambda_{IR}$ is taken over
by the infrared renormalization scale $\mu_{IR}$. Technically, the latter can be justified by the observation
that the object $\Delta I_{QR}(p^2)$ develops a pole for $d=3$ if evaluated in the heavy-baryon limit with
$M_R \to \infty $ \cite{Lutz:Kolomeitsev:2002,KSW}. Absorbing this pole into the counter terms of the chiral
Lagrangian introduces an ambiguity how to subtract the pole. The latter can be used to assign
$\bar I_{QR}(p^2)$ its natural chiral power, but, in addition to motivate the presence of the $\mu_{IR}$ term.

It should be possible to absorb the effect of $\delta I_{QR}$ into the bare parameters of the chiral Lagrangian.
This is confirmed by explicit calculations.  For the baryon octet masses at leading order the
relevant parameters are $b_{0}, b_D$ and $b_{F}$. We write
\begin{eqnarray}
b_{0,D,F}= b_{0,D,F}^{\chi{\rm-MS}}+\delta b_{0,D,F}\,, \qquad
d_{0,D}= d_{0,D}^{\chi{\rm-MS}}+\delta d_{0,D}\,.
\label{}
\end{eqnarray}
Making use of the explicit expressions of section three  we derive
\begin{eqnarray}
\delta b_0\,&=& \frac{\chiral{M}_{[8]}-2\,\mu_{IR}}{(4\pi f)^2}
\,\Big({\textstyle{13 \over 18}}\,D^2+{\textstyle{1\over 2}}\,F^2
+{\textstyle{7\over 18}}\,C^2\Big)
+ \mathcal{O} \left( \chiral{M}_{[10]}-\chiral{M}_{[8]}\right)\,,
\nonumber \\
\delta b_F&=&\frac{\chiral{M}_{[8]}-2\,\mu_{IR}}{(4 \pi f)^2}  \,
\Big({\textstyle{5\over 6}}\,D\,F + {\textstyle{5\over 36}}\,C^2\Big)
+ \mathcal{O} \left( \chiral{M}_{[10]}-\chiral{M}_{[8]}\right) \,, \qquad
\nonumber \\
\delta b_D &=& \frac{\chiral{M}_{[8]}-2\,\mu_{IR}}{(4 \pi f)^2} \, \Big(
{\textstyle{3\over 4}}\,F^2-{\textstyle{1\over 4}}\,D^2 -{\textstyle{1\over 6}}\,C^2\big)
+ \mathcal{O} \left( \chiral{M}_{[10]}-\chiral{M}_{[8]}\right)\,.
\label{result-delta-octet}
\end{eqnarray}
The results (\ref{result-delta-octet}) are instructive. They illustrate two
phenomena. The counter terms are proportional to $\chiral{M}_{[8]}- 2\,\mu_{IR} $. The
'$\chiral{M}_{[8]}$' term
is needed to guarantee $\bar I_{QR}(p^2) \sim Q$. The '$\mu_{IR}$' specifies the
running of the $Q^2$-terms on the infrared renormalization scale. It resembles the
cutoff dependence of the counter terms in the scheme of
\cite{Donoghue:Holstein:Borasoy:1999,Borasoy:Holstein:Lewis:Ouimet:2002}.
Using the values (\ref{large-Nc}) together with $\mu_{IR} \simeq 300$ MeV
we may estimate the importance of the $\mu_{IR}$-running. If (\ref{result-delta-octet}) is compared with
the typical tree-level values (\ref{tree-level-parameter}),  we conclude that the
$\mu_{IR}$-running is a crucial contribution to the counter terms. In fact, to arrive at baryon
masses that are independent on the infrared scale $\mu_{IR}$, the mass parameters in (\ref{chiral-L}) must
run as well:
\begin{eqnarray}
&& \chiral{M}_{[8]} \,\,=M+\frac{5}{3}\,\frac{\mu_{IR}}{8\pi^2 f^2}\,
\Delta^2 \,C^2 \,,
\nonumber\\
&&\chiral{M}_{[10]} =M+ \Delta+\frac{2}{3}\,\frac{\mu_{IR}}{8\pi^2 f^2}\,
\Delta^2 \,C^2 \,,
\label{mass-run}
\end{eqnarray}
where we introduce the scale invariant parameters $M$ and $\Delta$ (see also (\ref{octet-HB}, \ref{decuplet-HB})).
We do not detail here the role of $\delta I_R$. It was checked that the effect of the latter can
be absorbed into counter terms.

A similar analysis of the decuplet self energies is performed. For the baryon decuplet masses the
relevant parameters are $d_{0}$ and $d_{D}$. Applying the results
of section 4 the consequence of the finite renormalization $\delta I_{QR}$ is readily  worked
out. At leading order we obtain:
\begin{eqnarray}
\delta d_0\,&=&\frac{\chiral{M}_{[8]}-2\,\mu_{IR}}{(4\pi f)^2} \,\Big( {\textstyle{1 \over 12}}\,C^2+
{\textstyle{25\over 324}}\,H^2 \Big)+ \mathcal{O} \left( \chiral{M}_{[10]}-\chiral{M}_{[8]}\right)\,,
\nonumber\\
\delta d_D&=&\frac{\chiral{M}_{[8]}-2\,\mu_{IR}}{(4\pi f)^2} \,
\Big( {\textstyle{1\over 12}}\,C^2+{\textstyle{5\over 36}}\,H^2 \Big)+
\mathcal{O} \left( \chiral{M}_{[10]}-\chiral{M}_{[8]}\right)\,.
\label{result-delta-decuplet}
\end{eqnarray}
Again we conclude from (\ref{result-delta-decuplet}) that the $\mu_{IR}$-running
is a crucial element of the counter terms.

It is instructive to compare the $\chi$-MS scheme with other approaches. In the heavy-baryon
formulation the parameters, $b_0, b_F, b_D, d_0$ and $d_{D}$ that characterize the
order $Q^2$ counter terms are independent on the
ultraviolet renormalization scale $\mu_{\,UV}$. This implies for instance that in the absence of the decuplet
states $\delta I_Q$ can be absorbed
fully into counter terms of order $Q^4$ and higher. On the other
hand if we applied the $\overline{MS}$-scheme the $Q^2$-parameters would pick up
an ultraviolet scale dependence of the form
\begin{eqnarray}
b_{0,D,F}  \sim \frac{\chiral{M}_{[8]}}{(4 \pi f)^2}\,\ln \left(\frac{\chiral{M}_{[8]}}{\mu_{\,UV}} \right)\,,
\label{scale-MS}
\end{eqnarray}
which is a consequence of the $\overline{MS}$-scheme expression for $\delta I_R$.
Any renormalization scheme that restores the chiral power counting rules amounts to a
redefinition of counter terms in a manner that contributions of the form (\ref{scale-MS})
are hidden. Such terms are not expected from chiral counting rules.
In the presence of the decuplet state the $Q^2$-counter terms acquire a logarithmic running
on the ultraviolet renormalization scale. The latter is required to compensate for contributions
 proportional to the mesonic tadpole
$$\sim (\chiral{M}_{[10]}-\chiral{M}_{[8]} )\,I_Q\,.$$
All together we derive
\begin{eqnarray}
b^{\chi{\rm-MS}}_{0}\,&=& \frac{\mu_{IR}}{8\pi^2 f^2}
\,\Big({\textstyle{13 \over 18}}\,D^2+{\textstyle{1\over 2}}\,F^2
+{\textstyle{7\over 18}}\,C^2\Big)
+\frac{7}{36}\, \frac{\Delta}{(4\pi f)^2}\, C^2\,\ln (\mu^2_{\rm UV}) + \cdots
\,,
\nonumber \\
b^{\chi{\rm-MS}}_{F}&=&\frac{\mu_{IR}}{8\pi^2 f^2}  \,
\Big({\textstyle{5\over 6}}\,D\,F + {\textstyle{5\over 36}}\,C^2\Big)
+\frac{5}{72}\,\frac{\Delta}{(4 \pi f)^2} \, C^2 \,\ln (\mu^2_{\rm UV})+ \cdots
\,, \qquad
\nonumber \\
b^{\chi{\rm-MS}}_{D} &=&\frac{\mu_{IR}}{8\pi^2 f^2} \, \Big(
{\textstyle{3\over 4}}\,F^2-{\textstyle{1\over 4}}\,D^2 -{\textstyle{1\over 6}}\,C^2\big)
-\frac{1}{12}\,\frac{\Delta}{(4 \pi f)^2} \,C^2\,\ln (\mu^2_{\rm UV})+ \cdots
\,,
\nonumber\\
d^{\chi{\rm-MS}}_{0}\,&=&\frac{\mu_{IR}}{8\pi^2 f^2} \,\Big( {\textstyle{1 \over 12}}\,C^2+
{\textstyle{25\over 324}}\,H^2 \Big)
+\frac{1}{24}\,\frac{\Delta}{(4\pi f)^2} \,C^2\,\ln (\mu^2_{\rm UV})+ \cdots\,,
\nonumber\\
d^{\chi{\rm-MS}}_{D}&=&\frac{\mu_{IR}}{8\pi^2 f^2} \,
\Big( {\textstyle{1\over 12}}\,C^2+{\textstyle{5\over 36}}\,H^2 \Big)
+\frac{1}{24}\,\frac{\Delta}{(4\pi f)^2} \,C^2\,\ln (\mu^2_{\rm UV})+ \cdots\,,
\label{result-delta-running}
\end{eqnarray}
where we made explicit the scale dependence of the counter terms.

\newpage

\section{Results}

We begin with a discussion of the baryon-octet mass shifts. Applying to the
result (\ref{PV-octet}) the renormalization condition (\ref{def-scheme}, \ref{def-chiMS})
the results of the heavy-baryon formulation should be recovered upon a further expansion.
Indeed for the baryon octet with $B \in [8]$ we obtain
\begin{eqnarray}
&&\Delta M^{\rm loop}_{B \in [8]} =\sum_{Q\in [8], R\in [8]}
\left(\frac{m_Q}{4\,\pi\,f}\,G_{QR}^{(B)} \right)^2 \Bigg\{
\mu_{IR}   -\frac{\pi}{2}\,m_Q\Bigg\}
\nonumber\\
&& \quad +\sum_{Q\in [8], R\in [10]}
\left(\frac{m_Q}{4\,\pi\,f}\,G_{QR}^{(B)} \right)^2 \, \Bigg\{
\frac{\Delta }{3}\,\ln \frac{m_Q}{\mu_{\rm UV}}
+\frac{2}{3}\,\mu_{IR}\left(1-\frac{\Delta^2}{m_Q^2}\right)
 \nonumber\\
&& \quad
+\,\frac{1}{3}\,\left(1-\frac{\Delta^2}{m_Q^2}\right)\,\Bigg[
\sqrt{\Delta^2-m_Q^2}\,\ln \frac{\Delta + \sqrt{\Delta^2-m_Q^2}}{\Delta - \sqrt{\Delta^2-m_Q^2}}
- 2\,\Delta \, \ln \frac{M}{m_Q} \Bigg]\Bigg\}\,.
\label{octet-HB}
\end{eqnarray}
The coupling constants $G_{QR}^{(B)}$ in (\ref{octet-HB}) are
given in Tab. \ref{tab:octet-couling} in terms of $F,D,C$.
For the particular choice $\mu_{IR} \to 0$ we reproduce the results of
\cite{Bernard:Kaiser:Meissner:1993,Banerjee:Milana:1995,Lehnhart:Gegelia:Scherer:2005}. An additional term
proportional to
$m_Q^2\,\Delta $ from the decuplet intermediate states in \cite{Banerjee:Milana:1995} reflects
a slightly different renormalization scheme.
The effect of the latter can be generated  upon a finite renormalization of the $Q^2$-counter terms.
It is to be emphasized that the result (\ref{octet-HB}) is valid only for
\begin{eqnarray}
m_Q \sim \Delta \sim Q \,.
\label{HB-expansion}
\end{eqnarray}
Thus, one must not take the chiral limit of (\ref{octet-HB}) with $m_Q \to 0$ but $\Delta \neq 0$.
This fact reflects itself by the presence of the term $\sim \Delta^3\,\log \,m_Q$ in the last line
of (\ref{octet-HB}), i.e. the chiral limit would be logarithmically divergent.

\begin{figure}[t]
\begin{center}
\includegraphics[width=13cm,clip=true]{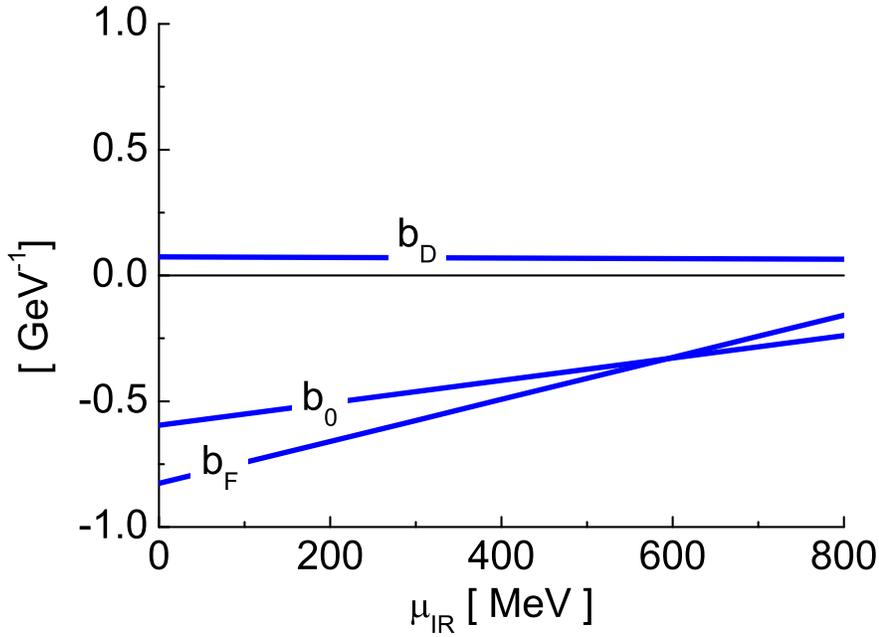}
\end{center}
\caption{It is shown the running of the $Q^2$ parameters $b_0, b_D$ and $b_F$ on the infrared scale
$\mu_{IR}$. We use $F=0.45, D=0.80$ and $C=0$ in (\ref{result-delta-running}). }
\label{fig:running}
\end{figure}

What happens with the octet masses once we include the loop correction?
At the given order the physical masses follow from (\ref{octet-HB}) by adding up the $Q^2$-terms
\begin{eqnarray}
M_B &=& \chiral{M}_{B}+\Delta M^{(2)}_{B} + \Delta M^{\rm loop}_{B}\,,
\label{def-loop}
\end{eqnarray}
where $\Delta M^{(2)}_{B}$ was given in (\ref{Q2-octet}).
The scale dependence of $\chiral{M}_{[8]} $
is specified in (\ref{mass-run}).
We first discuss the case where the decuplet intermediate state are omitted.
In the limit $C=0$ and  $\mu_{IR}=0$ the results of
\cite{Bernard:Kaiser:Meissner:1993,Lehnhart:Gegelia:Scherer:2005} are reproduced by  (\ref{octet-HB}).
The expressions of the heavy-baryon formulation of Jenkins and Manohar \cite{Bernard:Kaiser:Meissner:1993} and
EOMS scheme of the Mainz group \cite{Lehnhart:Gegelia:Scherer:2005} coincide. To illustrate the effect of the
infrared renormalization  scale we adjust the parameters $b_0,b_D,b_F$ and $M$ to obtain an optimal representation
of the octet masses and a pion-nucleon sigma term\footnote{We use the tree-level expressions
$$m_q\,\frac{d\,m_\pi^2}{d \,m_q} = m_\pi^2\,,\qquad m_q\,\frac{d\,m_K^2}{d \,m_q} = \frac{1}{2}\,m_\pi^2 \,,
\qquad m_q\,\frac{d\,m_\eta^2}{d \,m_q} = \frac{1}{3}\,m_\pi^2\,.$$}
\begin{eqnarray}
\sigma_{\pi N} = m_q\,\frac{d\,M_N}{d \,m_q} = 45 \,{\rm MeV} \,,
\label{def-sigma-pin}
\end{eqnarray}
for different choices of $\mu_{IR}$ \cite{Gasser:Leutwyler:Sainio:1991}. We remind the reader that we work in the isospin limit with degenerate
masses for the up- and down quarks, i.e. $m_u=m_d=m_q$.
The baryon masses and the sigma term
are manifestly scale independent once the $Q^2$ mass relation $m_\eta^2 = 4\,m_K^2/3 -m_\pi^2/3$ is used.
In contrast the parameters $b_0, b_D$ and $b_F$ are strongly scale dependent. As emphasized before
natural values can only be expected for appropriate choices of the renormalization scales.
The running of those parameters as determined by (\ref{result-delta-running}) is shown in
Fig. \ref{fig:running}. If insisting on $\mu_{IR}=0$ MeV the magnitudes of  the counter terms
are larger up to a factor of eight as compared to their tree-level values, which are
recalled in (\ref{tree-level-parameter}). As a consequence the chiral expansion appears poorly convergent.  This is
illustrated in Tab. \ref{tab:compare1}, where the baryon masses are split into their chiral moments for
the two choices $\mu_{IR}=0$ MeV and $\mu_{IR}=628$ MeV.  For the latter value the counter terms are
reasonably close to their tree-level values. In turn,  the loop corrections have natural size and there is
no longer an apparent problem with the convergence of the chiral expansion. However, there is still  a caveat.
The size of $\mu_{IR}$ for which we obtain naturally sized counter terms is somewhat large. In fact, as illustrated
in Fig. \ref{fig:running} even smaller counter terms are implied by a further increase with
$\mu_{IR} \simeq 800$ MeV.

\begin{table}[t]
\begin{center}
\begin{tabular}{|c|c|c|}\hline
 & (\ref{octet-HB}) with $ \mu_{IR}=0$\, MeV &  (\ref{octet-HB}) with $\mu_{IR}=628$\, MeV \\ \hline
$M_N$ [MeV] & $1018+\phantom{1}231-\phantom{1}307=\phantom{1}943 $ &  $1018-\phantom{11}44-\phantom{11}31=\phantom{1}943 $\\
$M_\Lambda$ [MeV] & $1018+\phantom{1}749-\phantom{1}655=1112 $ & $ 1018+\phantom{1}223-\phantom{1}129=1112 $\\
$M_\Sigma$ [MeV] & $1018+\phantom{1}839-\phantom{1}667=1189 $ & $ 1018+\phantom{1}303-\phantom{1}131=1189 $\\
$M_\Xi$ [MeV] & $1018+1311-1007=1322  $& $1018+\phantom{1}530-\phantom{1}226=1322$\\ \hline
\hline
&EOMS \cite{Lehnhart:Gegelia:Scherer:2005}&  LDR \cite{Donoghue:Holstein:1998}  \\ \hline
$M_N$ [MeV] & $1039+\phantom{1}240-\phantom{1}339=\phantom{1}940 $ & $ 1143-\phantom{1}237+\phantom{11}34=\phantom{1}940 $\\
$M_\Lambda$ [MeV] & $ 1039+\phantom{1}811-\phantom{1}737=1113 $&$ 1143-\phantom{11}86+\phantom{11}57=1114 $\\
$M_\Sigma$ [MeV] & $1039+\phantom{1}849-\phantom{1}696=1192 $& $ 1143-\phantom{111}5+\phantom{11}53=1191 $\\
$M_\Xi$ [MeV] & $1039+1400-1120=1319 $& $ 1143+\phantom{1}106+\phantom{11}77=1326 $\\ \hline
\end{tabular}
\caption{Baryon octet masses evaluated in the EOMS scheme\cite{Lehnhart:Gegelia:Scherer:2005} and
the cutoff scheme (LDR) \cite{Donoghue:Holstein:Borasoy:1999} as compared with (\ref{octet-HB}) evaluated for $C=0$.
The masses are decomposed into their chiral moments. }
\label{tab:compare1}
\end{center}
\end{table}

The octet masses within the 'no-decuplet scenario' are compared in Tab. \ref{tab:compare1} with the previous
studies \cite{Lehnhart:Gegelia:Scherer:2005} and \cite{Donoghue:Holstein:Borasoy:1999}. The choice $\mu_{IR}=0$ MeV
recovers the results of the heavy-baryon approach \cite{Lehnhart:Gegelia:Scherer:2005}.  Note the slightly
different values used for the parameters $F, D$ and $f$. It is interesting
to observe that the second choice with $\mu_{IR} = 628$ MeV resembles to some extent the results of the
cutoff-scheme suggested by Donoghue and Holstein \cite{Donoghue:Holstein:1998}. For small cutoff parameters,
$\Lambda_{DH}$, one may match
\begin{eqnarray}
\mu_{IR} = \frac{\pi}{2} \,\Lambda_{DH} \,.
\label{identify}
\end{eqnarray}
Given the identification (\ref{identify}) the cutoff dependence of the $Q^2$ parameter
obtained in \cite{Donoghue:Holstein:1998} is recovered with (\ref{result-delta-running}). Since the
study \cite{Donoghue:Holstein:Borasoy:1999} discusses the choice $\Lambda_{DH}= 400$ MeV in detail we
use the particular value $\mu_{IR} = 628$ MeV in Tab. \ref{tab:compare1} for a comparison. As is the case in
\cite{Donoghue:Holstein:Borasoy:1999} the $Q^3$ terms are reasonably small suggesting possibly a
converging expansion. There is, however, a striking difference of the two schemes. Whereas
Borasoy \cite{Borasoy:1999} obtains for $\Lambda_{DH}=400$ MeV a positive value for the strange-quark matrix
element of the nucleon\footnote{We use the tree-level expressions
$$m_s\,\frac{d\,m_\pi^2}{d \,m_s} = 0\,,\qquad m_s\,\frac{d\,m_K^2}{d \,m_s} = m_K^2-\frac{m_\pi^2}{2} \,,
\qquad m_s\,\frac{d\,m_\eta^2}{d \,m_s} = \frac{4}{3}\,\Big(m_K^2-\frac{m_\pi^2}{2}\Big)\,.$$},
\begin{eqnarray}
S_N= m_s \, \frac{d\,M_N}{d \,m_s} \,,
\label{def-sn}
\end{eqnarray}
with $S_N\simeq +163$ MeV, we derive $S_N\simeq  -273$ MeV. A negative value  is typically
obtained within the heavy baryon approach \cite{Lehnhart:Gegelia:Scherer:2005}. We will return to this
discrepancy in the course of including the decuplet degrees of freedom and when presenting the particular
summation scheme we are developing.

\begin{table}[t]
\begin{center}
\begin{tabular}{|c|c|c|}\hline
 & $\mu_{IR}=0$\, MeV &  $ \mu_{IR}=628$\, MeV \\ \hline
$M_N$ [MeV] & $1251+\phantom{11}89-\phantom{1}401=\phantom{1}939  $& $ 1565-\phantom{1}442-\phantom{1}184=\phantom{1}939 $\\
$M_\Lambda$ [MeV] & $1251+\phantom{1}789-\phantom{1}925=1116  $&$ 1565-\phantom{1}173-\phantom{1}276=1116 $\\
$M_\Sigma$ [MeV] & $1251+1376-1434=1193 $ & $1565-\phantom{11}76-\phantom{1}296=1193 $\\
$M_\Xi$ [MeV] & $1251+1783-1716=1318 $& $1565+\phantom{1}145-\phantom{1}391=1318 $\\
\hline \hline
$b_0\; \mathrm{[GeV^{-1}]}$ & $-1.38$ & $+0.07$\\
$b_D\; \mathrm{[GeV^{-1}]}$ & $+0.49$ & $+0.08$\\
$b_F\; \mathrm{[GeV^{-1}]}$ & $-0.93$ & $-0.32$\\
$\sigma_{\pi N}$ [MeV] & 45 & 45\\
$S_N$ [MeV] & $-708$ & $-708$\\ \hline
\end{tabular}
\caption{Baryon octet masses evaluated with (\ref{octet-HB}) for $C=1.6$ and $\mu_{UV}=800$ MeV.
The masses are decomposed into their chiral moments. }
\label{tab:HB-octet}
\end{center}
\end{table}

\begin{figure}[t]
\begin{center}
\includegraphics[width=13cm,clip=true]{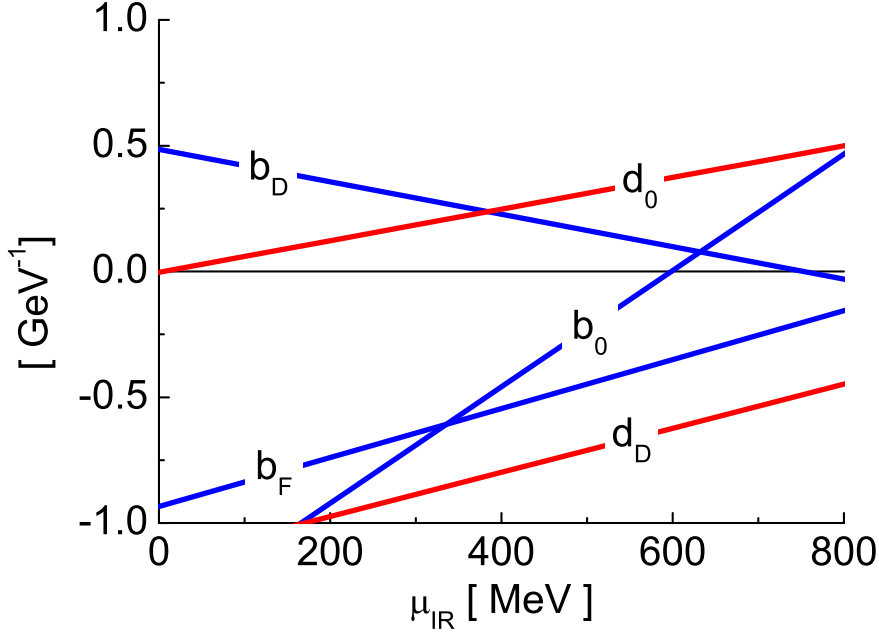}
\end{center}
\caption{It is shown the running of the $Q^2$ parameters $b_0, b_D, b_F$ and $d_0, d_D$ on the infrared scale
$\mu_{IR}$ at fixed $\mu_{UR}=800$ MeV.
We use $F=0.45, D=0.80$ and $C=1.60, H=1.65$ in (\ref{result-delta-running}). }
\label{fig:running-decuplet}
\end{figure}

We discuss the effect of the decuplet states. Since for $C\neq 0$ the additional
parameter, $\Delta$, is active in (\ref{octet-HB}) there is almost no predictive power for the
octet masses. For given values $F=0.45$, $D=0.80$ and $C=1.60$ together with
$f=92.4$ MeV it is possible to adjust the five parameters $b_0, b_D, b_F$ and $M, \Delta $ to
the four masses and the pion-nucleon sigma term $\sigma_{\pi N} = 45$ MeV exactly. We obtain
$\Delta\simeq 281$ MeV and $M \simeq 1251$ MeV manifestly independent on the ultraviolet and infrared
scales $\mu_{UV}$ and $\mu_{IR}$. The results of this analysis are collected in Tab. \ref{tab:HB-octet}
and in Fig. \ref{fig:running-decuplet}. Even though a perfect representation of the octet masses is achieved
the scenario is not convincing due to the considerably deteriorated convergence properties
of the expansion. This is in contrast to the cutoff scheme \cite{Donoghue:Holstein:1998} which claims good
convergence properties for the octet masses even if the decuplet intermediate states are
incorporated \cite{Borasoy:1999,Borasoy-et-al:2002}. The  value for the strange-quark matrix element of the
nucleon, $S_N$, varies from $+$168 MeV down to $+$53 MeV for the range of cutoff parameters
300 MeV $<\Lambda_{DH} <$ 600 MeV \cite{Borasoy:1999}. The number $S_N=-708$ MeV quoted in Tab. \ref{tab:HB-octet}
is in striking disagreement with the latter range.
It should be noted that the optimal values for $\Delta$ and $M$ depend sensitively on $F,D$ and $C$. It is
not always possible to reproduce the three mass splitting of the octet states exactly by a proper choice
of parameters. On the other hand a $\chi^2$ analysis reveals that its minimum is typically quite flat
in parameter space leaving some ambiguity how to fix the parameters. This ambiguity is
removed to a large extent once the decuplet masses are considered in addition.

Making use of the representation (\ref{PV-decuplet}) established in section 4 we
derive the loop correction of the decuplet states
\begin{eqnarray}
&&\Delta M^{\rm loop}_{B \in [10]} =\sum_{Q\in [8], R\in [8]}
\left(\frac{m_Q}{4\,\pi\,f}\,G_{QR}^{(B)} \right)^2 \, \Bigg\{
-\frac{\Delta }{6}\,\ln \frac{m_Q}{\mu_{\rm UV}}
+\frac{\mu_{IR}}{3}\,\left(1-\frac{\Delta^2}{m_Q^2}\right)
 \nonumber\\
&& \quad
+\,\frac{1}{6}\,\left(1-\frac{\Delta^2}{m_Q^2}\right)\,\Bigg[
\sqrt{\Delta^2-m_Q^2}\,\ln \frac{\Delta + \sqrt{\Delta^2-m_Q^2}}{\Delta - \sqrt{\Delta^2-m_Q^2}}
-2\,\Delta \, \ln \frac{M}{m_Q} \Bigg]\Bigg\}
\nonumber\\
&& \quad +
\sum_{Q\in [8], R\in [10]}
\left(\frac{m_Q}{4\,\pi\,f}\,G_{QR}^{(B)} \right)^2 \frac{5}{9}\,\Bigg\{
\mu_{IR}  -\frac{\pi}{2}\,m_Q\Bigg\}\,,
\label{decuplet-HB}
\end{eqnarray}
where a strict chiral expansion according to (\ref{HB-expansion}) is performed.
The coupling constants $G_{QR}^{(B)}$ in (\ref{decuplet-HB}) are
given in Tab. \ref{tab:decuplet-coupling} in terms of $C$ and $H$.
For $\mu_{IR} =0$ we recover the results of \cite{Banerjee}.
The total mass at order $Q^3$ follows with (\ref{def-loop}) and (\ref{mass-run}).
At this stage chiral symmetry is predictive. The parameters $\Delta\simeq 281$ MeV and $M \simeq 1251$ MeV
were set already to reproduce the baryon octet masses. We are left with $d_0$ and $d_D$ which
now determine the four masses of the decuplet states. A fit to the empirical masses
is summarized in Tab. \ref{tab:HB-decuplet} where we show results for two choices
of the infrared scale, $\mu_{IR}$, but at fixed $\mu_{UV}=800$ MeV. An amazingly accurate representation of
all decuplet masses is achieved. The running of the parameter $d_0$ and $d_D$ is shown in
Fig. \ref{fig:running-decuplet} together with the running of $b_0, b_D$ and $b_F$.
The counter terms appear 'most' naturally for 600 MeV $< \mu_{IR} <$ 800 MeV.
Incidentally, within that interval the parameters are within reach of the
large-$N_c$ expectations $b_D+b_F \sim d_D/3$ and
$d_0 \simeq b_0$. It is stressed that the latter range for $\mu_{IR}$ poses a problem since
we count $\mu_{IR} \sim Q$.

The deficiency of (\ref{decuplet-HB}) is most clearly unravelled by deriving the hadronic decay widths
of the decuplet states. Applying (\ref{HB-expansion}) to (\ref{PV-decuplet}) we obtain
\begin{eqnarray}
&&\Gamma^{\rm loop}_{B \in [10]} =\sum_{Q\in [8], R\in [8]}
\frac{\left(G_{QR}^{(B)}\right)^2}{24\,\pi\,f^2}\,\sqrt{\Delta^2-m_Q^2}^{\,3}\,\Theta \Big[\Delta-m_Q\Big]  \,.
\label{decuplet-HB-width}
\end{eqnarray}
Using the previously determined value $\Delta \simeq 281$ MeV we almost reproduce the decay width of the
$\Delta(1232)$ but overestimate the width of the $\Xi(1530)$ by about a factor of five (see Tab. \ref{tab:width}).
This is a disaster asking for a major reorganization of the expansion scheme.

\begin{table}[t]
\begin{center}
\begin{tabular}{|c|c|c|}\hline
& $\mu_{IR}=0$\, MeV &  $\mu_{IR}=628$\, MeV \\ \hline
$M_\Delta$ [MeV] &  $1532+\phantom{11}47-\phantom{1}347=1232 $&$ 1657-\phantom{1}376-\phantom{11}49=1232 $\\
$M_\Sigma$ [MeV] & $1532+\phantom{1}395-\phantom{1}544=1382 $&$ 1657-\phantom{1}195-\phantom{11}80=1382 $\\
$M_\Xi$ [MeV] & $1532+\phantom{1}742-\phantom{1}745=1529 $&$ 1657-\phantom{11}15-\phantom{1}114=1529 $\\
$M_\Omega$ [MeV] & $1532+1089-\phantom{1}949=1672 $&$ 1657+\phantom{1}166-\phantom{1}151=1672 $\\
\hline \hline
$d_0\; \mathrm{[GeV^{-1}]}$  & $-0.00$ & $+0.39$\\
$d_D\; \mathrm{[GeV^{-1}]}$ & $-1.15$ & $-0.60$\\ \hline
\end{tabular}
\caption{Baryon decuplet masses evaluated with (\ref{decuplet-HB}) for $C=1.60, H=1.65$ and $\mu_{UV}=800$ MeV.
The masses are decomposed into their chiral moments.}
\label{tab:HB-decuplet}
\end{center}
\end{table}

A natural summation approach is defined by performing a chiral loop expansion rather than a strict
chiral expansion: for a given truncation of the relativistic chiral Lagrangian
we take the loop expansion that is defined in  terms of the approximated Lagrangian seriously.
Clearly the number of loops we would consider is correlated with the chiral order to which the Lagrangian
is constructed. Also, a renormalization needs to be devised that installs the correct minimal chiral power of a
given loop function. The residual dependence on the renormalization scales is used to monitor the convergence
properties of the expansion and therewith to estimate the error encountered at a given truncation.

At the one loop order we are working here
the hadronic decay widths of the decuplet states are readily evaluated within the proposed scheme.
Taking the imaginary part of
(\ref{PV-decuplet}, \ref{decuplet-result-8}) in the limit $d=4$ we obtain without any expansion
\begin{eqnarray}
&&\Gamma^{\rm loop}_{B\in [10]} =\sum_{Q\in [8], R\in [8]}
\frac{\left(G_{QR}^{(B)}\right)^2}{24\,\pi\,f^2}\,\frac{M_R+E_R}{2\,M_B}\, p_{QR}^{\,3}  \,, \qquad
 E^2_R = M_R^2+p_{QR}^2 \,,
\label{decuplet-1-loop-width}
\end{eqnarray}
where the momenta, $p_{QR}$, were introduced already in (\ref{ipin-analytic}). A
conventional chiral expansion of (\ref{decuplet-1-loop-width}) reproduces the troublesome
result (\ref{decuplet-HB-width}). Taking (\ref{decuplet-1-loop-width}) face value the decay widths of the
decuplet states are collected in Tab. \ref{tab:width}. All but the isobar width are reproduced accurately.
The source of the striking difference lies in the phase-space factors $p_{QR}^3 $ and $(\Delta^2-m_Q^2)^{3/2}$.

\begin{table}[t]
\begin{center}
\begin{tabular}{|c||c|c|c|c|}
\hline
 &$\Gamma_\Delta$ [MeV] & $\Gamma_\Sigma $ [MeV] & $\Gamma_\Xi$ [MeV] & $\Gamma_\Omega $  [MeV] \\
\hline
\hline
Exp.   & 120 $\pm $ 5   & 36 $\pm $ 5    & 9.9 $\pm $ 1.9   &  0  \\
\hline
 (\ref{decuplet-HB-width})&  104 &  86 & 52 & 0 \\
\hline
  (\ref{decuplet-1-loop-width})&  73 &  34 & 12 & 0 \\
\hline
\end{tabular}
\caption{Total hadronic decay width of the baryon decuplet states. The widths follow
from (\ref{decuplet-HB-width}) and (\ref{decuplet-1-loop-width}) with $f=92.4$ MeV, $C = 1.6$
and $\Delta = 281$ MeV. }
\label{tab:width}
\end{center}
\end{table}

But how to perform a partial summation of the real part? If we use (\ref{decuplet-1-loop-width}) for
the hadronic width we must necessarily modify also the expression for the mass shifts. After all causality
relates the two. Here the Passarino-Veltman reduction offers a convenient and consistent method. The
mass shifts are expressed in terms of the renormalized objects $\bar I_{QR}(p^2)$ and $\bar I_Q$ introduced in
(\ref{def-master-loop}, \ref{def-scheme}, \ref{def-chiMS}). We stress that function $\bar I_{QR}(p^2)$
satisfies a once subtracted integral-dispersion representation as required by causality (see (\ref{disp-integral})).
Making use of the representation (\ref{PV-octet}, \ref{octet-result-8}, \ref{octet-result-10}) we obtain for
the octet states
\begin{eqnarray}
&&\Delta M^{\rm loop}_{B \in [8]} = \sum_{Q\in [8], R\in [8]}
\left(\frac{G_{QR}^{(B)}}{2\,f} \right)^2  \Bigg\{
\frac{M_R^2-M_B^2}{2\,M_B}\, \bar I_Q
\nonumber\\
&& \quad - \frac{(M_B+M_R)^2}{E_R+M_R}\, p^2_{QR}\,
\Big(\bar I_{QR}(M_B^2) + \frac{\bar I_Q}{M_R^2-m_Q^2}\,\Big)\Bigg\}\label{octet}
\\
&& \qquad \quad \;\,\,+\sum_{Q\in [8], R\in [10]}
\left(\frac{G_{QR}^{(B)}}{2\,f} \right)^2 \, \Bigg\{
\Bigg( \frac{(M_R-M_B)\,(M_R+M_B)^3+m_Q^4}{12\,M_B\,M^2_R}\,
\nonumber\\
&& \quad
- \frac{(Z\,(Z+2)-5)\,M_B^2+2\,(2\,Z\,(Z-1)-3)\,M_R\,M_B+2\,M_R^2}{12\,M_B\,M_R^2}\,m_Q^2\Bigg)\,
\bar I_Q
\nonumber\\
&& \quad  - \frac{2}{3}\,\frac{M_B^2}{M_R^2}\,\big(E_R+M_R\big)\,p_{QR}^{\,2}\,
\Big(\bar I_{QR}(M_B^2) + \frac{\bar I_Q}{M_R^2-m_Q^2}\Big) \Bigg\}
\,. \nonumber
\label{}
\end{eqnarray}
The previous result (\ref{octet-HB}) is recovered from (\ref{octet}) upon a chiral expansion. In contrast
to (\ref{octet-HB}) it is now possible to perform the chiral limit with $m_Q \to 0$. It follows
\begin{eqnarray}
&& \bar I_{QR} (p^2) \to \frac{\mu_{IR}}{8\,\pi^2\,\sqrt{p^2}}+
\frac{1}{(4\,\pi)^2}\,\frac{M_R^2-p^2}{2\,p^2}\,\log \frac{M_R^2-p^2}{M_R^2} \,,
\\
&& \bar I_Q \to 0 \,, \quad \qquad \qquad \qquad \qquad \qquad \qquad \qquad \qquad \qquad {\rm for} \qquad m_Q \to 0\,.
\nonumber
\label{chiral-limit}
\end{eqnarray}
The mesonic tadpole $\bar I_Q$ specified in (\ref{ren-quantities}) enjoys a logarithmic
dependence on the ultraviolet renormalization
scale $\mu_{UV}$ and the one-loop master function  $\bar I_{QR}(p^2)$ a linear dependence on the infrared
renormalization scale $\mu_{IR} $. By construction the result (\ref{octet}) is necessarily consistent with
all chiral Ward identities as discussed in section 5. The point is to avoid the poorly convergent expansion of
the coefficients in front of $\bar I_{QR}(p^2) $ and $\bar I_Q$.  This guarantees that the full imaginary
part of the self energy is recovered. The latter is proportional to $p_{QR}^3$ (see (\ref{disp-integral})).
We emphasize that (\ref{octet}) depends on the physical meson and baryon masses
$m_Q$ and $M_{R}$. This defines a self consistent summation since the masses of the intermediate baryon states in
(\ref{octet}) should match the total masses defined in (\ref{def-loop}). As long as the total masses are
sufficiently close to the physical ones it is clearly legitimate to use physical masses for the intermediate
states. Given the accuracy expected from Tabs. \ref{tab:HB-octet}-\ref{tab:HB-decuplet} this appears well justified
and we will do so in the following.

Like with all partial summation approaches there is a prize to pay. Either a scheme dependence or a
residual dependence on some renormalization scale remains. This is analogous to the residual cutoff
dependence of the scheme of Donoghue and Holstein \cite{Donoghue:Holstein:1998}. As long as such dependencies
are small and decreasing as higher order terms are included they do not pose a problem, rather, they offer
a convenient way to estimate the error encountered at a given truncation.

\begin{figure}[t]
\begin{center}
\includegraphics[width=13cm,clip=true]{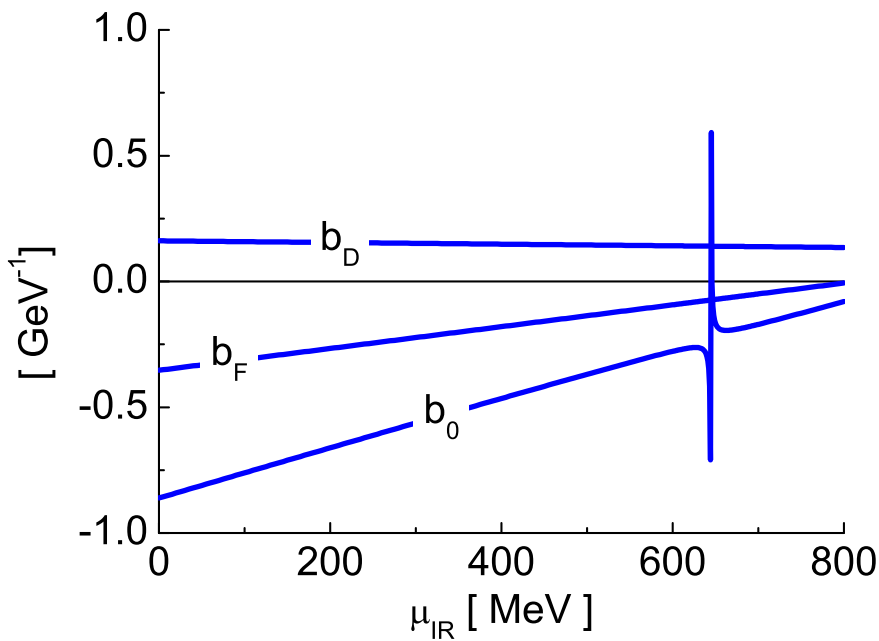}
\end{center}
\caption{The running of the $Q^2$ parameters $b_0, b_D$ and $b_F$ on the infrared scale
$\mu_{IR}$ is shown. The parameters are fitted so that (\ref{def-loop}) with
(\ref{Q2-octet}, \ref{octet}) reproduces the octet masses at fixed
$\sigma_{\pi N} =45$ MeV and $\mu_{UV}=800$ MeV. We use physical masses and $C=0$ in (\ref{octet}).  }
\label{fig:running-full}
\end{figure}

We discuss the implications of (\ref{octet}) first for $C=0$. The parameters $b_D, b_F$ are fitted to the mass
differences of the octet states. The parameter $b_0$ is used to obtain the  pion-nucleon sigma term,
$\sigma_{\pi N} = 45$ MeV and $\chiral{M}_{[8]}$ is set to recover the physical nucleon mass.
In Fig. \ref{fig:running-full} the $\mu_{IR}$ dependence of the $Q^2$ counter terms
that follow are shown. The pole-type structure in the running of the $b_0$ parameter at about
$\mu_{IR} \simeq 645$ MeV is striking, but a clear artifact of an unnatural large choice of $\mu_{IR}$.
Technically the pole arises since the self consistent evaluation of a sigma term requires a
matrix inversion: the algebraic equation
\begin{table}[t]
\begin{center}
\begin{tabular}{|c|c|c|}\hline
 & (\ref{octet}) with $ \mu_{IR}=0$\, MeV &  (\ref{octet}) with $\mu_{IR}=628$\, MeV \\ \hline
$M_N$ [MeV] & $\phantom{1}744+\phantom{1}400-\phantom{1}205=\phantom{1}939 $ &  $\phantom{1}845+\phantom{11}56+\phantom{11}37=\phantom{1}939 $\\
$M_\Lambda$ [MeV] & $\phantom{1}744+\phantom{1}671-\phantom{1}291=1125 $ & $\phantom{1}845+\phantom{11}87+\phantom{1}174=1106 $\\
$M_\Sigma$ [MeV] & $\phantom{1}744+\phantom{1}867-\phantom{1}411=1199 $ & $ \phantom{1}845+\phantom{1}257+\phantom{11}84=1187 $\\
$M_\Xi$ [MeV] & $\phantom{1}744+1040-\phantom{1}464=1320  $& $\phantom{1}845+\phantom{1}203+\phantom{1}268=1316$\\ \hline
\hline
& (\ref{octet}) with $ \mu_{IR}=300$\, MeV &  IR \cite{Ellis:Tang:1998}  \\ \hline
$M_N$ [MeV] & $\phantom{1}805+\phantom{1}223-\phantom{11}89=\phantom{1}939 $ & $ \phantom{1}733+\phantom{1}342-\phantom{1}160=\phantom{1}915 $\\
$M_\Lambda$ [MeV] & $\phantom{1}805+\phantom{1}380-\phantom{11}69=1116 $&$ \phantom{1}733+\phantom{1}671-\phantom{1}201=1204 $\\
$M_\Sigma$ [MeV] & $\phantom{1}805+\phantom{1}563-\phantom{1}175=1193 $& $ \phantom{1}733+\phantom{1}919-\phantom{1}494=1158 $\\
$M_\Xi$ [MeV] & $\phantom{1}805+\phantom{1}628-\phantom{1}115=1318 $& $ \phantom{1}733+1124-\phantom{1}589=1268 $\\ \hline
\end{tabular}
\caption{Baryon octet masses evaluated in the IR scheme \cite{Ellis:Tang:1998} and
as compared with (\ref{octet}) evaluated for $C=0$ and $\mu_{UV} = 800$ MeV.
The masses are decomposed into their chiral moments. }
\label{tab:compare2}
\end{center}
\end{table}
\begin{eqnarray}
m\,\frac{d\, M_B}{d \,m} &=& m\,\frac{d \,\Delta M_B^{(2)}}{d \,m} +
m\,\frac{d \,\Delta M^{\rm loop }_B}{d \,m}
\label{def-sigma-terms}\\
&=&m\,\frac{d \,\Delta M_B^{(2)}}{d \,m} + m\,\sum_R\,\frac{d\, M_R}{d \,m}
\frac{\partial \,\Delta M^{\rm loop }_B}{\partial \,M_R}
+ m\,\sum_Q\,\frac{d\, m_Q}{d \,m}
\frac{\partial\, \Delta M^{\rm loop }_B}{\partial \,m_Q}\,,
\nonumber
\end{eqnarray}
has to be solved for $d\, M /d \,m $. Here $m$ denotes a current quark mass, $m_Q$ the meson and
$M_{B}, M_R$ the baryon masses. Our computations of $\sigma_{\pi N}$ and $S_N$ are based on
tree-level expressions for $d\,m_Q /d\, m_q$ and $d\,m_Q / d\,m_s$ (see (\ref{def-sigma-pin}, \ref{def-sn})).
The determinant of
the system (\ref{def-sigma-terms})
\begin{eqnarray}
\det \Big[\delta_{BR }- \frac{\partial \,\Delta M^{\rm loop }_B}{\partial \,M_R}\Big]\,,
\label{det-sigma}
\end{eqnarray}
may vanish for a given choice of $\mu_{IR}$. In this case a
finite sigma term can be kept only  by an appropriately dialed infinite $b_0$ parameter. From (\ref{det-sigma}) it is
evident that at the one-loop level the determinant is a function of the infrared scale $\mu_{IR}$, the
physical masses and the parameter $F,D,C$
and $H$ only. This implies that the artifact at $\mu_{IR} \simeq 645$ MeV is to be removed by incorporating the
running of the latter parameters on $\mu_{IR}$. This should be done by performing a one-loop computation of the
hadronic vertex functions. Once this is done the pole structure in $b_0$ should disappear.
We claim, that without such a computation it is possible to study the
mass shifts in a reliable manner: if a range of $\mu_{IR}$ is selected where the expansion converges convincingly
such artifacts cannot occur. Our claim relies on the expectation that an evaluation of the vertex correction
is well converging within the same range of $\mu_{IR}$ where the mass expansion is well converging. If this is
true the tree-level values for the vertex should be a fair representation of the full vertex given a properly
selected range of $\mu_{IR}$. The consistency of this strategy will be addressed in a forthcoming
publication \cite{Semke:Lutz:2006}.

\begin{figure}[t]
\begin{center}
\includegraphics[width=13cm,clip=true]{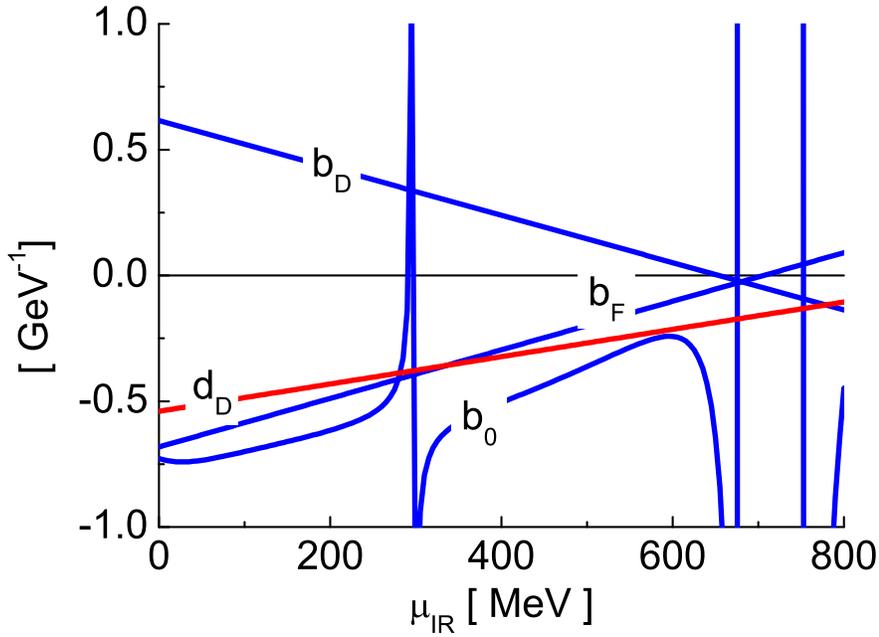}
\end{center}
\caption{The running of the $Q^2$ parameters $b_0, b_D, b_F$ and $d_D$ on the infrared scale
$\mu_{IR}$ is shown. The parameters are fitted so that (\ref{def-loop}) with
(\ref{Q2-octet}, \ref{octet}, \ref{decuplet}) reproduces the baryon masses at fixed
$\sigma_{\pi N} =45$ MeV and $\mu_{UV}=800$ MeV. We use physical masses and $C=1.6 $ in
(\ref{octet}) and assume $d_0=b_0$.  }
\label{fig:running-full-decuplet}
\end{figure}

If Fig. \ref{fig:running-full} is compared to Fig. \ref{fig:running} a reduction of the
$Q^2$ parameters within the natural window 200 MeV $< \mu_{IR}<$ 400 MeV is observed, in particular the
magnitude of $b_F$ is reduced significantly. As a consequence
the chiral loop expansion appears much better convergent within that interval. This is demonstrated in
Tab. \ref{tab:compare2} where the result of this analysis is shown for different values of the infrared
scale $\mu_{IR}$. It is important to realize that for $\mu_{IR}$ close to the troublesome value
$\mu_{IR} \simeq 645$ MeV, the convergence properties are not acceptable. To illustrate this the results
for the particular choice $\mu_{IR}=628$ MeV are included in Tab. \ref{tab:compare2}. The latter value
was used previously in Tab. \ref{tab:compare1} in order to offer a fair comparison with the results of the
cutoff scheme of Donoghue and Holstein \cite{Donoghue:Holstein:1998}.
The partial summation defined by (\ref{octet}) leads to convincing convergence properties
within the natural window 200 MeV $< \mu_{IR}<$ 400 MeV. This is a significant improvement over
the results displayed in Tab. \ref{tab:compare1} based on a strict chiral expansion (\ref{octet-HB}).
In order to estimate the uncertainty at the given truncation we provide the variance of
$\chiral{M}_{[8]}$ and $S_N$ in the selected window. We obtain  with
785 MeV $< \,\chiral{M}_{[8]}\,\,<$ 823 MeV and 27 MeV $< S_N< $ 70 MeV a reasonably small variation.
These ranges are not realistic yet. They will be modified once we incorporate the decuplet fields
into the analysis.

The partial summation (\ref{octet}) is compared to the one suggested by  Becher and Leutwyler
\cite{Becher:Leutwyler:1999}. The computation of Ellis and Torikoshi \cite{Ellis:Torikoshi:1999}, that applied the
IR scheme \cite{Becher:Leutwyler:1999}, are quoted in Tab. \ref{tab:compare2}. Those results are clearly
much less proving. It should be noted that Ellis and Torikoshi sacrificed somewhat the quality of their fit
by insisting on $S_N=200$ MeV. Their best fit suggested $S_N=360$ MeV insisting on $\sigma_{\pi N}= 45$ MeV.

We switch on the fluctuation of the baryon-octet states into virtual meson baryon-decuplet pairs, i.e.
we use $C=1.60$ in (\ref{octet}).
Given only (\ref{octet}) it is not possible to determine the parameters $b_0$  from
the octet masses and a given pion-nucleon sigma term. The former depends on the values of $d_0$ and $d_D$ via
the dependence of the octet masses on the masses of the decuplet states. Thus a consistent analysis requires
a simultaneous evaluation of the decuplet masses.  Applying the representation
(\ref{PV-decuplet}, \ref{decuplet-result-8}, \ref{decuplet-result-10}) it is straightforward to perform
the summation for the decuplet that is analogous to (\ref{octet}). In contrast
to the mass shifts of the octet states (\ref{octet}) we obtain a result that is independent on
the parameter $Z$:
\begin{eqnarray}
&&\Delta M^{\rm loop}_{B\in [10]} = \sum_{Q\in [8], R\in [8]}
\left(\frac{G_{QR}^{(B)}}{2\,f} \right)^2  \Bigg\{
\Bigg( \frac{(M_R-M_B)\,(M_R+M_B)^3+m_Q^4}{24\,M^3_B}\,
\nonumber\\
&& \quad
- \frac{3\,M_B^2+2\,M_R\,M_B+2\,M_R^2}{24\,M^3_B}\,m_Q^2\Bigg)\,
\bar I_Q
\nonumber\\
&& \quad -\frac{1}{3}\,\big( E_R +M_R\big)\,p_{QR}^{\,2}\,
\Big(\bar I_{QR}(M_B^2)+ \frac{\bar I_Q}{M_R^2-m_Q^2}\Big) \Bigg\}
\label{decuplet}
\\
&& \qquad \quad \;\,\,+\sum_{Q\in [8], R\in [10]}
\left(\frac{G_{QR}^{(B)}}{2\,f} \right)^2 \, \Bigg\{ \Bigg( \frac{(M_B+M_R)^2\,m_Q^4}{36\,M_B^3\,M_R^2}
\nonumber\\
&&\quad
+\frac{3\,M_B^4-2\,M^3_B\,M_R+3\,M_B^2\,M_R^2-2\,M_R^4}{36\,M_B^3\,M_R^2}
\,m_Q^2
\nonumber\\
&&\quad +\frac{M_R^4+M_B^4+12\,M_R^2\,M_B^2-2\,M_R\,M_B\,(M_B^2+M_R^2)}{36\,M^3_B\,M^2_R}\,
(M^2_R-M^2_B)
\Bigg)\,\bar I_Q
\nonumber\\
&& \quad  -\frac{(M_B+M_R)^2}{9\,M_R^2}\,\frac{2\,E_R\,(E_R-M_R)+5\,M_R^2}{E_R+M_R}\,
p_{QR}^{\,2}\,\Big(\bar I_{QR}(M_B^2)+ \frac{\bar I_Q}{M_R^2-m_Q^2}\Big)  \Bigg\}\,.
\nonumber
\end{eqnarray}

The result of a combined fit of the baryon octet and decuplet masses as implied by
(\ref{octet}, \ref{decuplet}) is shown in Fig. \ref{fig:running-full-decuplet}. In the
analysis the pion-nucleon sigma term is kept at $\sigma_{\pi N} = 45$ MeV and
in addition we assumed the large-$N_c$ relation $d_0=b_0$.
The infrared running of the parameters is complicated with
pole structures in $b_0$ around $\mu_{IR} \simeq 295$ MeV and $\mu_{IR} \simeq 660$ MeV.
As discussed above such pole structures do not cause a problem, they rather help to identify
the natural window. They suggest the natural window: 350 MeV $<\mu_{IR}<$ 550 MeV.
Only within that range we may expect good convergence properties and therefore
can trust the expansion scheme. Like we already observed within a strict chiral expansion scheme, the
inclusion of the decuplet states shifts the required range of $\mu_{IR}$ to somewhat higher masses.

\begin{table}[t]
\begin{center}
\begin{tabular}{|c|c|c|}\hline
 & $\mu_{IR}=350$\, MeV &  $ \mu_{IR}=550$\, MeV \\ \hline
$M_N$ [MeV] & $1057+\phantom{111}3-\phantom{1}121=\phantom{1}939  $& $ \phantom{1}969+\phantom{11}59-\phantom{11}89=\phantom{1}939 $\\
$M_\Lambda$ [MeV] & $1057+\phantom{1}229-\phantom{1}150=1136  $&$ \phantom{1}969+\phantom{1}166-\phantom{111}9=1126 $\\
$M_\Sigma$ [MeV] & $1057+\phantom{1}575-\phantom{1}425=1207 $ & $\phantom{1}969+\phantom{1}285-\phantom{11}54=1200 $\\
$M_\Xi$ [MeV] & $1057+\phantom{1}627-\phantom{1}361=1323 $& $\phantom{1}969+\phantom{1}322+\phantom{11}19=1320 $\\
\hline \hline
$M_\Delta$ [MeV] & $\phantom{1}713+\phantom{1}610-\phantom{11}91=1232  $& $ \phantom{1}919+\phantom{1}301+\phantom{11}12=1232 $\\
$M_\Sigma$ [MeV] & $\phantom{1}713+\phantom{1}716-\phantom{11}49=1380  $&$ \phantom{1}919+\phantom{1}374+\phantom{11}83=1376 $\\
$M_\Xi$ [MeV] & $\phantom{1}713+\phantom{1}821-\phantom{111}4=1530 $ & $\phantom{1}919+\phantom{1}447+\phantom{1}160=1526 $\\
$M_\Omega$ [MeV] & $\phantom{1}713+\phantom{1}927+\phantom{11}34=1674 $& $\phantom{1}919+\phantom{1}520+\phantom{1}235=1674 $\\
\hline \hline
$b_0\; \mathrm{[GeV^{-1}]}$ & $-0.58$ & $-0.29$\\
$b_D\; \mathrm{[GeV^{-1}]}$ & $+0.29$ & $+0.10$\\
$b_F\; \mathrm{[GeV^{-1}]}$ & $-0.34$ & $-0.15$\\
$d_0\; \mathrm{[GeV^{-1}]}$ & $-0.58$ & $-0.29$\\
$d_D\; \mathrm{[GeV^{-1}]}$ & $-0.35$ & $-0.24$\\
$\sigma_{\pi N}$ [MeV] & 45 & 45\\
$S_N$ [MeV] & $-28$ & $-98$\\ \hline
\end{tabular}
\caption{The parameters are fitted so that (\ref{def-loop}) with
(\ref{Q2-octet}, \ref{octet}) reproduces the octet masses at fixed
$\sigma_{\pi N} =45$ MeV. We use $C=1.6, Z=0.72$ and $\mu_{UV}=800$ MeV.
The masses are decomposed into their chiral moments. }
\label{tab:full-decuplet}
\end{center}
\end{table}

In Tab. \ref{tab:full-decuplet} a detailed summary of the results is compiled for two
choices of the infrared scale. For $\mu_{IR}=550$ MeV an amazingly consistent expansion pattern
arises. The lower value $\mu_{IR}=350$ MeV may be acceptable but has less convincing convergence
properties. It is reassuring that for the larger value the parameters are quite consistent with
the large-$N_c$ expectation $b_F+b_D \sim d_D/3$. Since we assumed the large-$N_c$ result, $d_0=b_0$,
in the fits this is an important consistency check. Slight variations ($d_0/b_0=1.0 \pm 0.2$) around the latter
assumption change our prediction for $S_N$ by less than 1 MeV.

\section{Summary}

We evaluated the baryon octet and decuplet self energies at the one-loop level applying
the covariant chiral Lagrangian. It is argued that an ambiguity persists within the $\chi$-MS scheme
on how to restore the chiral counting rules. This leads to the presence of an infrared
renormalization scale that can be used to optimize the speed of convergence. Performing a
strict chiral expansion the physical parameters are independent on the infrared scale.
However, the size of the $Q^2$ counter terms depend on this scale. In turn the apparent
convergence properties reflect the choice of that scale. Insisting on a reasonable
range the convergence properties of the chiral expansion are improved considerably, though
not yet reaching a convincing state.

We discussed in detail the octet and decuplet mass shifts as they arise in a strict chiral
expansion. All masses can be reproduced accurately by a fit of the $Q^2$ counter terms.
Good convergence properties require, however, an anomalously large infrared scale.
The deficiency of this strategy is most clearly visible when studying the
hadronic decay widths of the decuplet states. Here the conventional expansion fails miserably
at leading order.

A summation approach was defined by performing a chiral loop expansion rather than a strict
chiral expansion: for a given truncation of the relativistic chiral Lagrangian
we take the loop expansion that is defined in  terms of the approximated Lagrangian seriously.
The number of loops we would consider is correlated with the chiral order to which the Lagrangian
is constructed. A renormalization based on the Passarino-Veltman reduction was devised that
installs the correct minimal chiral power of a
given loop function. The residual dependence on the renormalization scales is used to monitor the convergence
properties of the expansion and therewith estimate the error encountered at a given truncation.
Within the proposed scheme the hadronic decay widths are recovered reasonably well. In addition
the octet and decuplet masses can be reproduced accurately with a small residual dependence on the
infrared renormalization scale only. Good convergence properties are found for natural
values of the infrared scale. Based on our analysis we predict

\begin{eqnarray}
S_N = m_s \,\frac{d \,M_N}{d\,m_s} = (-63\pm 35)\,{\rm MeV} \nonumber \,.
\end{eqnarray}
for a given pion-nucleon sigma term $\sigma_{\pi N}=45 $ MeV.

\section*{Acknowledgments}

M.F.M.L. acknowledges fruitful discussions with B. Friman, E.E. Kolomeitsev and A. Schwenk.

\newpage

\section{Appendix}

We investigate an arbitrary one-loop integral involving any number of scalar propagators and open
Lorentz indices that arises when computing one-baryon processes. It is proven that
it is sufficient to renormalize the scalar master-loop functions of the
Passarino-Veltman reduction in a manner that the latter are compatible with the expectation
of chiral counting rules. The method of complete induction is applied.

Consider the generic integral
\begin{eqnarray}
&&I_{k,r}^{\mu_1 \cdots \mu_n}(q,P)=i\,(-1)^{k+r+1}\,\mu^{4-d} \,\int \frac{d^d l}{(2\pi)^d}\,
l^{\mu_1} \cdots  l^{\mu_n}\,S_{k,r}(l,q,P)\,,
\nonumber\\
&& S_{k,r}(l,\,q,\,P)=\left( \Pi_{i=1}^k\frac{1}{(l+q_i)^2-m_i^2}\right)
\left( \Pi_{j=1}^r \frac{1}{(l+p_j)^2-M_j^2}\right)\,,
\label{def-tensor-integral}
\end{eqnarray}
where we assume
\begin{eqnarray}
&& p^\mu_i = P^\mu + q^\mu_{k+i} \,, \qquad q^\mu_i \sim Q \,,\qquad
 P^\mu \sim Q^0 \,, \qquad P^2-M_i^2  \sim Q \,,
 \nonumber\\
&& m^2_i  \sim Q^2\,,\qquad M_i\sim Q^0\,,\qquad M_i-M_j \sim Q^2\,.
\end{eqnarray}
The typical 4-momentum of the initial and final heavy particle is denoted
with $P_\mu$. A naive application of power counting
rules suggests
\begin{eqnarray}
I_{k,r}^{\mu_1 \cdots \mu_n}(q,P) \sim Q^{4+n-2\,k-r} \,.
\label{goal}
\end{eqnarray}
It is to be shown that the renormalized integral (\ref{def-tensor-integral})
is compatible  with (\ref{goal}) provided (\ref{goal}) holds for the special case $n=0$.
In the following we assume that the renormalization procedure commutes with the
Passarino-Veltman reduction and that $I_{k,r}(q,p)$  is renormalized for arbitrary $k\geq0$
and $r\geq 0$ in a manner that the counting rule (\ref{goal}) is realized.

It is convenient to reformulate the claim (\ref{goal}). For this purpose we consider the class of tensors,
$T_{\mu_1 \cdots \mu_n}(q,p)$, that is
constructed as the product of the 4-momenta $q_i,\,p_j$, and the metric tensors $g^{\mu_i \mu_j}$.
We claim that it is sufficient to proof that the projection of $I_{k,r}^{\mu_1\cdots \mu_n}(q,P)$ with
any of the tensors $T_{\mu_1 \cdots \mu_n}(q,p)$ scales as expected
\begin{eqnarray}
T_{\mu_1 \cdots \mu_n}(q,p)\, I_{kr}^{\mu_1\cdots \mu_n}(q,P) \sim Q^{4+n-2\,k-r+\#(q)} \,,
\label{reformulate:goal}
\end{eqnarray}
where $\#(q) $ is the number of small momenta, $q$, involved in the tensor.
Clearly, if (\ref{goal}) is valid (\ref{reformulate:goal}) is an immediate consequence. The reverse conclusion
is less trivial. Suppose (\ref{reformulate:goal}) holds but (\ref{goal}) is not true. If we can lead this
to a contradiction we are done. The latter would imply that the tensor integral has a contribution orthogonal
to all possible tensors $T_{\mu_1 \cdots \mu_n}(q,p)$. This cannot be: applying the Feynman parametrization
of the loop integral it follows that the tensor, $I_{k,r}^{\mu_1\cdots \mu_n}(q,P)$, must
be composed out of the 4-momenta $q_i, \,p_i $ and the metric tensor.

The proof is prepared further by two  observations. First for the integrals with $r=0$,
\begin{eqnarray}
I_{k,0}^{\mu_1 \cdots \mu_n}(q,P) \sim Q^{4+n-2\,k} \,,
\label{r-zero-case}
\end{eqnarray}
there is nothing to be done: the counting rule is observed trivially. Since there is only one typical
scale involved, $q_i\sim m_i \sim Q$, dimensional counting
is preserved: the integrals are independent on $M_i$ and $P_\mu$. The
standard observation that dimensional regularization
introduces the renormalization scale in a logarithmic manner and therefore preserves
dimensional counting is recalled.

The second observation concerns the special case $k=0$. Before renormalization the
counting rule
\begin{eqnarray}
I_{0,r}^{\mu_1 \cdots \mu_n}(q,P) \;\slash{\sim } \;Q^{4+n-r} \,,
\label{k-zero-case}
\end{eqnarray}
is violated. For instance for $n=0$ the tadpole
integral should scale at least with $Q^3$. Since it depends on one heavy mass parameter, $M$,
this can be achieved only after renormalization: it must be put to zero:
\begin{eqnarray}
I_{0,1} (q,P)\,\Big|_R = 0 \,.
\label{ren-tadpole}
\end{eqnarray}
This requirement has important consequences. Consider any scalar integral with $r=r_0>1$. It may be Taylor expanded
in the 4-momenta $q^\mu_i$ around the point $q^\mu_i =0$. The assumption $q^2_i \ll M^2_j $ implies that
the integrals are real, which guarantees that the expansion converges in the kinematical domain studied.
Any coefficient can be expressed in terms of the objects $I_{0, 1}(0,P)$. Thus, if the
renormalization is to commute with the
Passarino-Veltman reduction and also with the Taylor expansion around $q^\mu_i =0$ we must insist on
\begin{eqnarray}
I_{0,r} (q,P)\,\Big|_R = 0 \,, \qquad \forall \quad r \,.
\label{k-zero-a}
\end{eqnarray}
Owing to the Passarino-Veltman reduction it follows
\begin{eqnarray}
I_{0,r}^{\mu_1 \cdots \mu_n}(q,P) \Big|_R =0 \,, \qquad \forall \quad r \,\& \, n \,,
\label{k-zero-b}
\end{eqnarray}
which is compatible with the minimal expected dimensional power $4+n-r$.

We turn to the proof of the main claim of this Appendix. It is constructed by
induction in the countable parameter $k+r$. For $k+r=1$ the
counting rule (\ref{goal}) is observed evidently. The cases $(k,r)=(1,0)$ and $(k,r)=(0,1)$ are
included in (\ref{r-zero-case}) and (\ref{k-zero-b}).
We proceed with the induction part of the proof, i.e. we assume that our claim is valid for
$k+r \leq m$ but any positive integer number for $n$. Based on this assumption we have to
show that the counting rule (\ref{goal}) is
realized for the renormalized loop functions characterized by $k+r=m+1$ and any $n$.

Our task is solved by an additional induction in $n$, i.e. we assume that (\ref{goal}) holds
for $k+r\leq m+1$ and n, but show that it holds also for $ k+r = m+1$ and $n+1$. The particular case
$n=0$ is our primary assumption that all scalar master loop functions are renormalized
compatibly with (\ref{goal}). Due to
(\ref{k-zero-b}) it is sufficient to consider the cases with $k\geq 1$. Thus we may perform a
change of variables in the integral
\begin{eqnarray}
 \tilde l = l+q_1 \,, \qquad \tilde q_i = q_i-q_1 \,, \qquad \tilde p_j= p_j-q_1\,.
 \label{shift}
\end{eqnarray}
The task is reduced to the proof that
(\ref{goal}) holds for $I^{\mu_1 \cdots \mu_{n+1}}_{k,r}(\tilde q,P)$ with $\tilde q_1=0$ and
$k+r= m+1$. According
to (\ref{reformulate:goal}) it is sufficient to consider contractions with tensors build out of
the 4-momenta $\tilde q_i, \tilde p_j$ and the metric tensor. Given a particular tensor we
distinguish three different cases: a) only products of metric tensors are involved, b) at least
one momentum $\tilde q_i$ is involved, c) at least one momentum $\tilde p_j$ is involved.

\begin{itemize}
\item[a)]
We may write
\begin{eqnarray}
T_{\mu_1 \cdots \mu_{n+1}} (\tilde q,\tilde p) = g_{\mu_1 \mu_2}\,T_{\mu_3 \cdots \mu_{n+1}} (\tilde q,\tilde p)\,,
\label{case-a}
\end{eqnarray}
where we assumed without loss of generality that the first two indices are connected via the
metric tensor. It is sufficient to show that
\begin{eqnarray}
g_{\mu_1 \mu_2}\,I_{k,r}^{\mu_1 \cdots \mu_{n+1}}(\tilde q,P) \sim Q^{4+n+1-2\,k-r}\,,
\label{to-be-shown-a}
\end{eqnarray}
holds. Due to $\tilde q_1 =0$ this is readily achieved. Consider the manipulation implied by the identity
$\tilde l^2 = (\tilde l^2-m_1^2)+m_1^2$. We obtain
\begin{eqnarray}
g_{\mu_1 \mu_2}\,I_{k,r}^{\mu_1 \cdots \mu_{n+1}}(\tilde q,P) = m^2_1\,I_{k,r}^{\mu_3 \cdots \mu_{n+1}}(\tilde q,P)
-I_{k-1,r}^{\mu_3 \cdots \mu_{n+1}}(\tilde q ,P)\,,
\label{result-a}
\end{eqnarray}
where the second term of (\ref{result-a}) involves $k-1$ soft propagators only. Thus by assumption
(\ref{to-be-shown-a}) follows.

\item[b)]

We may write
\begin{eqnarray}
T_{\mu_1 \cdots \mu_{n+1}} (\tilde q,\tilde p) = (\tilde q_{x})_{\mu_1}\,T_{\mu_2  \cdots  \mu_{n+1}} (\tilde q,\tilde p)\,,
\label{case-b}
\end{eqnarray}
where  $1< x\leq k$.  It is sufficient to prove
\begin{eqnarray}
(\tilde q_{x})_{\mu_1}\,I_{k,r}^{\mu_1 \cdots \mu_{n+1}}(\tilde q,P) \sim Q^{4+n+2-2\,k-r}\,,
\label{to-be-shown-b}
\end{eqnarray}
which follows applying the identity
\begin{eqnarray}
2\,\tilde l \cdot \tilde q_x = ((\tilde l+\tilde q_x)^2-m_x^2)-(\tilde l^2-m_1^2) +m_x^2-m_1^2-q_x^2\,.
\end{eqnarray}
It holds
\begin{eqnarray}
&& (\tilde q_{x})_{\mu_1}\,I_{k,r}^{\mu_1 \cdots \mu_{n+1}}(\tilde q,P) =
(m_x^2-m_1^2-\tilde q_x^2)\,I_{k,r}^{\mu_2 \cdots \mu_{n+1}}(\tilde q,P)
\nonumber\\
&& \qquad -I_{k-1,r}^{\mu_2 \cdots  \mu_{n+1}}(\tilde q_{ \bot x},P)
+I_{k-1,r}^{\mu_2 \cdots  \mu_{n+1}}(\tilde q_{\bot  1},P)
\,,
\label{result-b}
\end{eqnarray}
where we apply the notation
\begin{eqnarray}
q_{\bot x} = \big\{q_1, \cdots ,\hat q_x ,\cdots ,q_{r+k}\big\} \,.
\label{notation-def}
\end{eqnarray}
The notation (\ref{notation-def}) is sufficient in (\ref{result-b}) since the correlation
of a particular mass parameter $m_i$ with the momentum $q_i$ or $\tilde q_i$ as introduced
in (\ref{def-tensor-integral}) is untouched. From (\ref{result-b}) the desired property
(\ref{to-be-shown-b}) is immediate, given the induction assumptions.

\item[c)]

We may write
\begin{eqnarray}
T_{\mu_1 \cdots \mu_{n+1}} (\tilde q,\tilde p) = (\tilde p_{x })_{\mu_1}\,T_{\mu_2 \cdots   \mu_{n+1}} (\tilde q,\tilde p)\,,
\label{case-b}
\end{eqnarray}
where  $1\leq  x\leq r$.
It is sufficient to derive the scaling law
\begin{eqnarray}
(\tilde p_{x})_{\mu_1}\,I_{k,r}^{\mu_1 \cdots \mu_{n+1}}(\tilde q,P) \sim Q^{4+n+1-2\,k-r}\,.
\label{to-be-shown-c}
\end{eqnarray}
The latter can be derived by applying the identity
\begin{eqnarray}
2\,\tilde l \cdot \tilde p_x = ((\tilde l+\tilde p_x)^2-M_x^2)-(\tilde l^2-m_1^2) +M_x^2-m_1^2-p_x^2\,.
\end{eqnarray}
It holds
\begin{eqnarray}
&& (\tilde p_{x})_{\mu_1}\,I_{k,r}^{\mu_1 \cdots \mu_{n+1}}(\tilde q,P) =
(M_x^2-m_1^2-\tilde p_x^2)\,I_{k,r}^{\mu_2 \cdots \mu_{n+1}}(\tilde q,P)
\nonumber\\
&& \qquad -I_{k-1,r}^{\mu_2 \cdots  \mu_{n+1}}(\tilde q_{ \bot r+x},P)
+I_{k-1,r}^{\mu_2 \cdots  \mu_{n+1}}(\tilde q_{\bot r+ 1},P)
\,.
\label{result-c}
\end{eqnarray}
From (\ref{result-c}) the desired property
(\ref{to-be-shown-c}) follows given the induction assumptions.

\end{itemize}

The proof is completed.


\newpage






\begin{thebibliography}{9}
 %\bibitem[*]{ifj}
%\bibitem[\dagger]{humboldt}
%\bibitem[\ddagger]{ml}

\bibitem{Jenkins:1992}
E. Jenkins, Nucl. Phys. {\bf B 368} (1992) 190.

\bibitem{Bernard:Kaiser:Meissner:1993}
V. Bernard, N. Kaiser and U.-G. Mei\ss ner, Z. Phys. {\bf C 60} (1993) 111.

\bibitem{Borasoy:Meissner:1997}
B. Borasoy, U.-G. Mei\ss ner, Ann. Phys. {\bf 254} (1997) 192.


\bibitem{Ellis:Torikoshi:1999}
P.J. Ellis and K. Torikoshi, Phys. Rev. {\bf C 61} (1999) 015205.

\bibitem{Lehnhart:Gegelia:Scherer:2005}
B.C. Lehnhart, J. Gegelia and S. Scherer, J. Phys. {\bf G 31}(2005) 89.

\bibitem{Gasser:Sainio:Svarc:1988}
J. Gasser, M.E. Sainio and A. Svarc, Nucl. Phys. {\bf B 307} (1988) 779.

\bibitem{Gasser:Leutwyler:1984}
J. Gasser and H. Leutwyler, Nucl. Phys. {\bf B 250} (1985) 465.

\bibitem{Jenkins:Manohar:1991}
E. Jenkins and A. Manohar, Phys. Lett. {\bf B 255} (1991) 558.

\bibitem{Donoghue:Holstein:1998}
J.F. Donoghue and B.R. Holstein, Phys. Lett. {\bf B 436} (1998) 331.

\bibitem{Donoghue:Holstein:Borasoy:1999}
J.F. Donoghue, B.R. Holstein and B. Borasoy, Phys. Rev. {\bf D 59} (1999) 036002.

\bibitem{Borasoy:Holstein:Lewis:Ouimet:2002}
B. Borasoy et al., Phys. Rev. {\bf D 66} (2002) 094020.

\bibitem{Ellis:Tang:1998}
P.J. Ellis and H.-B. Tang, Phys. Rev. {\bf C 57} (1998)  3356.

\bibitem{Becher:Leutwyler:1999}
T. Becher and H. Leutwyler, Eur. Phys. J. {\bf C 9} (1999) 643.

\bibitem{Lutz:2000}
M. Lutz, Nucl. Phys. {\bf A 677} (2000) 241.

\bibitem{Gegelia:Japaridze:1999}
J. Gegelia and G. Japaridze, Phys. Rev. {\bf D 60} (1999) 114038.

\bibitem{Fuchs:Gegelia:Japaridze:Scherer:2003}
T. Fuchs et al., Phys. Rev. {\bf D 68} (2003) 056005.

\bibitem{Lutz:Kolomeitsev:2002}
M.F.M. Lutz and E. E. Kolomeitsev, Nucl. Phys. {\bf A 700} (2002) 193.

\bibitem{Hacker:Wies:Gegelia:Scherer:2005}
C. Hacker, N. Wies, J. Gegelia and S. Scherer, Phys. Rev. {\bf C 72} (2005) 055203.


\bibitem{Bernard:Hemmert:Meissner:2005}
V. Bernard, T. R. Hemmert, U.-G. Mei\ss ner, Phys. Lett. {\bf B 622} (2005) 141.

\bibitem{Passarino:Veltman:1979}
G. Passarino and M. Veltman, Nucl. Phys. {\bf B 160} (1979) 151.


\bibitem{Krause:1990}
A. Krause, Helv. Phys. Acta {\bf 63} (1990) 3.

\bibitem{Bernard:Hemmert:Meissner:2003}
V. Bernard, T.R. Hemmert and U.-G. Mei\ss ner, Phys. Lett. {\bf B 565} (2003) 137.


\bibitem{Dashen}
R. F. Dashen, E. Jenkins and A.V. Manohar, Phys. Rev. {\bf D 51} (1995) 3697.

\bibitem{Jenkins}
E. Jenkins, Phys. Rev. {\bf D 53} (1996) 2625.

\bibitem{Jenkins:Manohar}
E. Jenkins and A. Manohar, Phys. Lett. {\bf B 259} (1991) 353.

\bibitem{Okun}
L.B. Okun, {\it Leptons and Quarks}, Amsterdam, North-Holland (1982).

\bibitem{Kolomeitsev:Lutz:2004}
E.E. Kolomeitsev and M.F.M. Lutz,
Phys. Lett. {\bf B 585} (2004) 243.

\bibitem{Lutz:Korpa:2002}
M.F.M. Lutz and C.L. Korpa, Nucl. Phys. {\bf A 700} (2002) 309.

\bibitem{Banerjee:Milana:1995}
M.K. Banerjee and J. Milana, Phys. Rev. {\bf D 52} (1995) 6451.


\bibitem{KSW}
D.B. Kaplan, M.J. Savage and M.B. Wise, Nucl. Phys. {\bf B 534} (1998) 329.

\bibitem{Gasser:Leutwyler:Sainio:1991}
J. Gasser, H. Leutwyler, M.E. Sainio, Phys. Lett. {\bf B 253} (1991) 252.

\bibitem{Borasoy:1999}
B. Borasoy, Eur. Phys. J. {\bf C 8} (1999) 121.

\bibitem{Borasoy-et-al:2002}
B. Borasoy et al., Phys. Rev. {\bf D 66} (2003) 094020.

\bibitem{Banerjee}
M.K. Banerjee and  J. Milana, Phys. Rev. {\bf D 54} (1996) 5804.


\bibitem{Semke:Lutz:2006}
A. Semke and M.F.M. Lutz, in preparation.




\end{thebibliography}
\end{document}